\documentclass[11pt]{article}
\usepackage{amssymb}
\makeatletter\@addtoreset {equation}{section}\makeatother

\newtheorem{theorem}{Theorem}
\newtheorem{lemma}{Lemma}
\newtheorem{corollary}{Corollary}
\newtheorem{proposition}{Proposition}

\newtheorem{remark}{Remark}

\newenvironment{proof}{
    \noindent {\it Proof.}}{\hfill$\Box$
}
\newenvironment{proof1}{
    \noindent {\it Proof }}{\hfill$\Box$
}

\usepackage[dvips]{epsfig}
\usepackage{graphicx}

\begin{document}

\title{\bf Periodic oscillations of dark solitons \\ in parabolic potentials}
\author{Dmitry E. Pelinovsky$^{\dagger}$ and Panayotis G. Kevrekidis$^{\dagger  \dagger}$ \\
{\small $^{\dagger}$ Department of Mathematics, McMaster
University, Hamilton, Ontario, Canada, L8S 4K1} \\
{\small $^{\dagger \dagger}$ Department of Mathematics and
Statistics, University of Massachusetts, Amherst, MA 01003 } }
\date{\today}
\maketitle

\begin{abstract}
We reformulate the Gross--Pitaevskii equation with an external
parabolic potential as a discrete dynamical system, by using the
basis of Hermite functions. We consider small amplitude stationary
solutions with a single node, called dark solitons, and examine
their existence and linear stability. Furthermore, we prove the
persistence of a periodic motion in a neighborhood of such
solutions. Our results are corroborated by numerical computations
elucidating the existence, linear stability and dynamics of the
relevant solutions.
\end{abstract}

\section{Introduction}

We address the Gross-Pitaevskii (GP) equation with an external
parabolic potential
\begin{equation}
\label{GP} i U_T = - \frac{1}{2} U_{XX} + \epsilon^2 X^2 U +
\sigma |U|^2 U,
\end{equation}
where $U(X,T) : \mathbb{R} \times \mathbb{R}_+ \mapsto \mathbb{C}$
is decaying to zero as $|x| \to \infty$, $\epsilon \in \mathbb{R}$
is the strength of the external potential and $\sigma = 1$
($\sigma = -1$) is normalized for the defocusing (focusing) cubic
nonlinearity. This equation is of particular interest in the
context of Bose-Einstein condensates, i.e., dilute alkali vapors
at near-zero temperatures, where dynamics of localized dips in the
ground state trapped by the magnetically induced, confining
potential $V(X) = \epsilon^2 X^2$ is studied in many recent
papers, see review in \cite{Konotop}. A question of particular
interest concerns whether the localized density dips oscillate
periodically near the center point $X = 0$ of the potential
$V(X)$. If the motion of a localized dip is truly periodic, the
frequency of periodic oscillations is to be found \cite{Kon05},
while if the periodic oscillations are destroyed due to emission
of radiation, the gradual change in the amplitude of oscillations
is to be followed for sufficiently small $\epsilon$ \cite{PFK05}.
Numerical simulations show radiation and amplitude changes if the
confining parabolic potential is perturbed by a periodic potential
while no radiation and time-periodic oscillations in the case of
purely parabolic confinement \cite{proukakis}.

If $\sigma = 1$, a localized dip on the ground state of the GP
equation (\ref{GP}) in the formal limit $\epsilon \to 0$
represents the so-called dark soliton of the defocusing nonlinear
Schr\"{o}dinger (NLS) equation, which is the reason why we use the
term ''dark soliton" for a localized solution. Persistence and
stability of a dark soliton of the defocusing NLS equation in the
presence of an exponentially decaying potential $V(X)$ was studied
in our previous paper \cite{PK07}, where methods of
Lyapunov--Schmidt reductions, Evans functions and the stability
theory in Pontryagin space were employed. These methods can not be
applied to the potential $V(X) = \epsilon^2 X^2$ since the
potential deforms drastically the spectrum of the linearized
problem: the continuous spectral band at $\epsilon = 0$ becomes an
infinite sequence of isolated eigenvalues for $\epsilon \neq 0$.
Therefore, we do not use here the limit $\epsilon \to 0$.
Moreover, we transform the GP equation (\ref{GP}) to the
$\epsilon$-independent form
\begin{equation}
\label{GP-zero-epsilon} i u_t = - \frac{1}{2} u_{xx} + \frac{1}{2}
x^2 u + \sigma |u|^2 u,
\end{equation}
by the scaling transformation $x = \lambda X$, $t = \lambda^2 T$,
and $u(x,t) = \lambda^{-1} U(X,T)$ with $\lambda = 2^{1/4}
\epsilon^{1/2}$.

Substitution $u(x,t) = e^{-\frac{i}{2} t - i \mu t} \phi(x)$
reduces equation (\ref{GP-zero-epsilon}) to the second-order
non-autonomous ODE
\begin{equation}
\label{ODE} - \frac{1}{2} \phi''(x) + \frac{1}{2} x^2 \phi(x) +
\sigma \phi^3(x) = \left( \mu + \frac{1}{2} \right) \phi(x),
\end{equation}
where $\phi : \mathbb{R} \mapsto \mathbb{R}$. A strong solution of
the ODE (\ref{ODE}) is {\em said} to be a dark soliton if
$\phi(x)$ is odd on $x \in \mathbb{R}$, has no zeros on $x \in
\mathbb{R}_+$, and decays to zero sufficiently fast as $|x| \to
\infty$. A classification of all localized solutions of the
second-order ODE (\ref{ODE}) and their construction with a
rigorous shooting method is suggested in recent work
\cite{Alfimov}.

Substitution of $u(x,t) = e^{-\frac{i}{2} t - i \mu t} \left[ \phi(x)
+ v(x,t) + i w(x,t) \right]$ reduces equation
(\ref{GP-zero-epsilon}) to the PDE system
\begin{eqnarray}
\label{PDE} \left\{ \begin{array}{rcl} v_t & = & {\cal L}_- w + 2
\sigma \phi(x) v w + \sigma (v^2 + w^2) w, \\
-w_t & = & {\cal L}_+ v + \sigma \phi(x) (3 v^2 + w^2) + \sigma
(v^2 + w^2) v, \end{array} \right.
\end{eqnarray}
where $(v,w) : \mathbb{R} \times \mathbb{R}_+ \mapsto
\mathbb{R}^2$ and ${\cal L}_{\pm}$ are self-adjoint
Schr\"{o}dinger operators in $L^2(\mathbb{R})$
\begin{equation}
\label{operators} \left\{ \begin{array}{rcl} {\cal L}_+ & = & -
\frac{1}{2}
\partial_x^2 + \frac{1}{2} x^2 - \frac{1}{2} - \mu + 3 \sigma
\phi^2(x), \\ {\cal L}_- & = & - \frac{1}{2} \partial_x^2 +
\frac{1}{2} x^2 - \frac{1}{2} - \mu + \sigma \phi^2(x).
\end{array} \right.
\end{equation}
Solutions of the PDE (\ref{GP-zero-epsilon}) are considered in
space
\begin{equation}
\label{function-space-1} {\cal H}_1(\mathbb{R}) = \{ u \in
H^1(\mathbb{R}) : \;\; x u \in L^2(\mathbb{R}) \}
\end{equation}
equipped with the norm
\begin{equation}
\label{Sobolev-space-norm} \| u \|^2_{{\cal H}_1} =
\int_{\mathbb{R}} \left( |u'(x)|^2 + (x^2 + 1) |u(x)|^2 \right)
dx.
\end{equation}
Similarly, the domain of operators ${\cal L}_{\pm}$ in
(\ref{operators}) is defined in space
\begin{equation}
\label{function-space-2} {\cal H}_2(\mathbb{R}) = \{ u \in
H^2(\mathbb{R}) : \;\; x^2 u \in L^2(\mathbb{R}) \}.
\end{equation}

The PDE system (\ref{PDE}) is a Hamiltonian system with the
standard symplectic structure and the Hamiltonian function in the
form
\begin{equation}
\label{Ham-function} H = \frac{1}{2} \left( v,{\cal L}_+ v \right)
+ \frac{1}{2} \left( w,{\cal L}_-w \right) + \sigma (\phi v, v^2 +
w^2) + \frac{\sigma}{4} (v^2+w^2,v^2+w^2),
\end{equation}
where $(\cdot,\cdot)$ denotes a standard inner product in
$L^2(\mathbb{R})$. The Hamiltonian function $H$ is bounded if $v,w
\in {\cal H}_1(\mathbb{R})$ and is constant in time $t$. Due to
the gauge invariance of the PDE (\ref{GP-zero-epsilon}), there
exists an additional quantity
\begin{equation}
\label{gauge-fun} Q = 2 (\phi,v) + (v,v) + (w,w),
\end{equation}
which is constant in time $t$. Global existence of solutions of
the initial-value problem associated with the PDE
(\ref{GP-zero-epsilon}) in space $u \in {\cal H}_1(\mathbb{R})$
for all $t \in \mathbb{R}_+$ has been proved (see Proposition 2.2
in \cite{Carles}).

Considering the linear part of the PDE system (\ref{PDE}), one can
separate the variables in the form $v(x,t) = v(x) e^{\lambda t}$,
$w(x,t) = w(x) e^{\lambda t}$ and obtain the linear problem
\begin{equation}
\label{spectrum} {\cal L}_+ v = - \lambda w, \quad {\cal L}_- w =
\lambda v
\end{equation}
for the spectral parameter $\lambda \in \mathbb{C}$ and the
eigenvector $(v,w) \in L^2(\mathbb{R},\mathbb{C}^2)$. Fix $\mu \in
\mathbb{R}$ such that a stationary solution of the ODE (\ref{ODE})
exists and $\phi \in {\cal H}_1(\mathbb{R})$. Then, the linear
problem (\ref{spectrum}) admits an exact solution
\begin{equation}
\label{exact-solution-spectrum} \lambda = \pm i : \qquad v =
\phi'(x), \quad w = \mp i x \phi(x),
\end{equation}
and $(v,w) \in L^2(\mathbb{R},\mathbb{C}^2)$. Additionally, for
any $\mu \in \mathbb{R}$, for which the stationary solution
$\phi(x)$ is smooth with respect to parameter $\mu$, the linear
problem (\ref{spectrum}) admits another exact solution for zero
eigenvalue $\lambda = 0$ of geometric multiplicity one and
algebraic multiplicity two:
\begin{equation}
\label{exact-solution-zero} {\cal L}_- \phi(x) = 0, \qquad {\cal
L}_+ \partial_{\mu} \phi(x) = - \phi(x).
\end{equation}

The main part of this work is devoted to the study of periodic
solutions of the PDE system (\ref{PDE}) for values of $\mu$ near
$\mu = 1$. This special value corresponds to the second eigenvalue
of the linear Schr\"{o}dinger operator
\begin{equation}
\label{L} {\cal L} = - \frac{1}{2} \partial_x^2 + \frac{1}{2} x^2
- \frac{1}{2}
\end{equation}
with the eigenfunction $\phi(x) = \varepsilon x e^{-x^2/2}$. Here
the parameter $\varepsilon$ is arbitrary in the linear problem and
it parameterizes the corresponding family of stationary solutions
$(\mu,\phi(x))$ of the nonlinear ODE (\ref{ODE}), which bifurcates
from the small-amplitude eigenmode \cite{Alfimov}.

A periodic solution of the PDE system (\ref{PDE}) bifurcates from
the linear eigenmodes $v(x) = \delta e^{i \tau} \phi'(x)$ and
$w(x) = \mp i \delta e^{i \tau} x \phi(x)$ corresponding to the
eigenvalue pair $\lambda = \pm i$ of the linear problem
(\ref{spectrum}). Here $\delta$ and $\tau$ are two real-valued
parameters, which are arbitrary in the linear problem and
parameterize the corresponding family of periodic solutions
$(v,w)$ of the PDE system (\ref{PDE}). An additional parameter
$\alpha$ comes from the projection of the solution $(v,w)$ to the
geometric kernel of the linear problem (\ref{spectrum}) with the
eigenmode $v(x) = 0$ and $w(x) = \alpha \phi(x)$. Using this
construction, the main result of our paper is described by the
following theorem.

\begin{theorem}
\label{theorem-main} Let $\varepsilon$ and $\delta$ be
sufficiently small and let $\alpha,\tau$ be arbitrary. There
exists a unique family of solutions of the ODE (\ref{ODE}) such
that
\begin{equation}
\label{property-1} \| \phi - \varepsilon x e^{-x^2/2} \|_{{\cal
H}_1} \leq C_1 \varepsilon^3, \quad \left| \mu - 1 - \frac{3
\sigma \varepsilon^2}{\sqrt{32 \pi}} \right| \leq C_2
\varepsilon^4,
\end{equation}
for some $\varepsilon$-independent constants $C_1,C_2 > 0$. There
exists a family of time-periodic space-localized solutions of the
PDE system (\ref{PDE}) such that $(v,w) \in {\cal
H}_1(\mathbb{R},\mathbb{R}^2)$ for all $t \in \mathbb{R}$,
\begin{equation}
\label{property-2} v\left(x,t + \frac{2\pi}{\Omega}\right) =
v(x,t), \quad w\left(x,t + \frac{2\pi}{\Omega}\right) = w(x,t),
\qquad \forall (x,t) \in \mathbb{R}^2,
\end{equation}
with the bounds
\begin{eqnarray}
\label{property-3} \left\| v(\cdot,t) - \delta \phi'(x)
\cos(\Omega t + \tau) \right\|_{{\cal H}_1}
& \leq & C_3 \varepsilon \delta^2, \\
\label{property-4} \left\| w(\cdot,t) - \delta \left[ x \phi(x)
\sin(\Omega t + \tau) + \alpha \phi(x) \right] \right\|_{{\cal
H}_1} & \leq & C_4 \varepsilon \delta^2,
\end{eqnarray}
and $|\Omega - 1 | \leq C_5 \varepsilon^2 \delta^2$ for some
($\varepsilon,\delta$)-independent constants $C_3,C_4,C_5 > 0$.
\end{theorem}

The periodic solution of Theorem \ref{theorem-main} has four free
parameters $(\varepsilon,\delta,\tau,\alpha)$ which are associated
with projections to the four eigenmodes
(\ref{exact-solution-spectrum}) and (\ref{exact-solution-zero}) of
the linear problem (\ref{spectrum}). Parameters $\tau$ and
$\alpha$ can be set to zero due to two obvious symmetries of the
PDE (\ref{GP-zero-epsilon}): the gauge transformation $u(x,t)
\mapsto u(x,t) e^{i \alpha}$, $\forall \alpha \in \mathbb{R}$ and
the reversibility transformation $u(x,t) \mapsto \bar{u}(x,-t)$,
$\forall t \in \mathbb{R}$.

Although the eigenmodes (\ref{exact-solution-spectrum}) and
(\ref{exact-solution-zero}) persist for all $\mu \in \mathbb{R}$,
existence of periodic orbits of the GP equation
(\ref{GP-zero-epsilon}) is only proved near $\mu = 1$. This is due
to the fact that the non-resonance conditions $n \neq {\rm Im}
\lambda_m$, $\forall n,m \in \mathbb{N}$ are proved to be
satisfied only in this domain, where $\lambda_m$ denote other
isolated eigenvalues of the linear problem (\ref{spectrum}) with
${\rm Im} \lambda_m > 0$ which are different from $\lambda = i$.
Thus, resonances do not occur near the value $\mu = 1$. The
construction of the periodic orbit is complicated due to the
existence of translational eigenmodes associated with the double
zero eigenvalue $\lambda = 0$ of the linear problem
(\ref{spectrum}).

Our main result is in agreement with Theorem 2.1 in \cite{Gus},
where the Newton's law of particle dynamics is obtained in a more
general context of multi-dimensional confining potentials and
general nonlinear functions of the GP equation $i \dot{\psi} = -
\nabla^2 \psi + V(x) \psi - f(\psi)$. The Newton's law is derived
for parameters $(a,p)$ of the solitary wave solution of the
unperturbed equation with $V(x) \equiv 0$ and it takes the form
\begin{equation}
\label{Newton-law} \dot{a} = 2 p, \qquad \dot{p} = - \nabla V(a).
\end{equation}
Adopting our notations for the time variable and the potential
function of the GP equation (\ref{GP-zero-epsilon}), we rewrite
the Newton's law (\ref{Newton-law}) in the explicit form $\ddot{a}
+ a = 0$, which recovers the frequency $\Omega = 1$ of the
periodic solution of Theorem \ref{theorem-main} in the linear
approximation $\delta \to 0$.

There are several differences between results of Theorem 2.1 in
\cite{Gus} and our Theorem \ref{theorem-main}. First, the Newton's
law (\ref{Newton-law}) is valid on finite time intervals and in
the limit when the localization length of the stationary solution
$\phi(x)$ is much smaller than the confinement length of the
potential $V(x)$. This situation corresponds to the original GP
equation (\ref{GP}) in the limit $\epsilon \to 0$. Second, the
exact periodicity is not guaranteed by the periodic solutions of
the Newton's law (\ref{Newton-law}) because of the remainder
terms. Lastly, the frequency $\Omega = 1$ of the Newton's law is
independent of the nonlinear function $f(\psi)$ and the nonlinear
corrections in $(a,p)$. In our case, the result of Theorem
\ref{theorem-main} is valid for all time intervals, the exact
periodicity is guaranteed, and the frequency $\Omega$ changes with
parameters $\delta$. On the other hand, our results are valid in
the limit $\mu \to 1$, which is far from the limit $\epsilon \to
0$ of the GP equation (\ref{GP}).

Note that the oscillations of the dark solitons in the GP equation
(\ref{GP-zero-epsilon}) with the frequency $\Omega = 1$ were
predicted from the Ehrenfest Theorem in much earlier works (see
references in \cite{Kon05} and \cite{Gus}). However, it was argued
that this frequency is not observed in numerical simulations of
the original GP equation (\ref{GP}) with $\sigma = 1$ for
sufficiently small $\epsilon$ \cite{Kon05,PFK05,proukakis}. It was
suggested in these works (see review in \cite{Konotop}) that dark
solitons oscillate with a smaller frequency $\Omega =
\frac{1}{\sqrt{2}}$. We will show that both frequencies occur in
the spectrum of the linear problem (\ref{spectrum}) in the
corresponding limit but the non-resonance conditions are not
satisfied for either frequency in this limit.

Our strategy for the proof of Theorem \ref{theorem-main} is to use
a complete set of Hermite functions and to reformulate the
evolution problem for the PDE (\ref{GP-zero-epsilon}) as an
infinite-dimensional discrete dynamical system for coefficients of
the decomposition (Section 2). Existence of stationary solutions
$\phi(x)$ of the ODE (\ref{ODE}) and spectral stability of
stationary solutions in the linear problem (\ref{spectrum}) are
studied in the framework of the discrete dynamical system (Section
3). The proof of existence of periodic solutions of the PDE system
(\ref{PDE}) relies on construction of periodic orbits in the
discrete dynamical system (Section 4). The analytical results are
verified with numerical approximations of solutions of the ODE
(\ref{ODE}), eigenvalues of the linear problem (\ref{spectrum})
and solutions of the GP equation (\ref{GP-zero-epsilon}) (Section
5). Distribution of eigenvalues of the linear problem
(\ref{spectrum}) in the limit $\mu \to \infty$ for $\sigma = 1$ is
also analyzed with formal asymptotic methods (Appendix A).

\section{Formalism of the discrete dynamical system}

The set of Hermite functions is defined by the standard
expressions \cite{AS}:
\begin{equation}
\label{Hermite-function} \phi_n(x) = \frac{1}{\sqrt{2^n n!
\sqrt{\pi}}} H_n(x) e^{-x^2/2}, \qquad \forall n = 0,1,2,3,...,
\end{equation}
where $H_n(x)$ denote the Hermite polynomials, e.g. $H_0 = 1$,
$H_1 = 2x$, $H_2 = 4 x^2 - 2$, $H_3 = 8 x^3 - 12 x$, etc. Since
the Hermite functions are eigenfunctions of the linear
Schr\"{o}dinger equation
\begin{equation}
\label{Schrodinger-Hermite} -\frac{1}{2} \phi_n''(x) + \frac{1}{2}
x^2 \phi_n(x) =  \left( n + \frac{1}{2} \right) \phi_n(x), \qquad
\forall n = 0,1,2,3,...,
\end{equation}
the Sturm--Liouville theory implies that the set of Hermite
functions $\{ \phi_n(x) \}_{n = 0}^{\infty}$ forms an orthogonal
basis in $L^2(\mathbb{R})$. The normalization coefficients in the
expressions (\ref{Hermite-function}) ensure that the Hermite
functions have unit $L^2$-norm, such that
\begin{equation}
\label{orthogonality} (\phi_n,\phi_m) = \delta_{n,m}, \qquad
\forall n,m = 0,1,2,3,....
\end{equation}
We represent a solution $u(x,t)$ of the GP equation
(\ref{GP-zero-epsilon}) by the series of eigenfunctions
\begin{equation}
\label{solution-series} u(x,t) = e^{-\frac{i}{2} t} \sum_{n =
0}^{\infty} a_n(t) \phi_n(x)
\end{equation}
where the components $(a_0,a_1,a_2,...)$ form a vector ${\bf a}$
on $\mathbb{N}$. When the series representation
(\ref{solution-series}) is substituted to the GP equation
(\ref{GP-zero-epsilon}), the PDE problem is converted to the
discrete dynamical system
\begin{equation}
\label{discrete-system} i \dot{a}_n = n a_n + \sigma
\sum_{(n_1,n_2,n_3)} K_{n,n_1,n_2,n_3} a_{n_1} \bar{a}_{n_2}
a_{n_3},  \qquad \forall n = 0,1,2,3,...,
\end{equation}
where $K_{n,n_1,n_2,n_3} = (\phi_n,\phi_{n_1} \phi_{n_2}
\phi_{n_3})$. We shall use a convention to avoid specifying the
range of non-negative integers $(n_1,n_2,n_3)$ and $n$ in the
summation signs of the dynamical system (\ref{discrete-system}).
Let $l^2_s(\mathbb{N})$ be a weighted discrete $l^2$-space
equipped with the standard norm
\begin{equation}
\label{discrete-norm} \| {\bf a} \|^2_{l^2_s} = \sum_{n =
0}^{\infty} (1 + n)^{2s} |a_n|^2 < \infty, \quad \forall s \in
\mathbb{R}.
\end{equation}
Since the set $\{ \phi_n(x) \}_{n = 0}^{\infty}$ forms an
orthonormal basis in $L^2(\mathbb{R})$, we note the isometry $\| u
\|^2_{L^2} = \| {\bf a}\|^2_{l^2}$, so that $u \in
L^2(\mathbb{R})$ if and only if ${\bf a} \in l^2(\mathbb{N})$. On
the other hand, we need an equivalence between the space ${\cal
H}_1(\mathbb{R})$ for the function $u(x)$ and the space
$l^2_s(\mathbb{N})$ for the vector ${\bf a}$. In addition, we need
to determine the domain and range of the vector field of the
discrete dynamical system (\ref{discrete-system}). These results
are described in Lemmas \ref{lemma-embedding} and
\ref{lemma-phase-space}.

\begin{lemma}
Let $u(x) = \sum\limits_{m = 0}^{\infty} a_n \phi_n(x)$. Then $u
\in {\cal H}_1(\mathbb{R})$ if and only if ${\bf a} \in
l^2_{1/2}(\mathbb{N})$. \label{lemma-embedding}
\end{lemma}

\begin{proof}
It follows directly that
\begin{eqnarray*}
\| u \|^2_{{\cal H}_1} & = &  \int_{\mathbb{R}} \left(
|u'(x)|^2 + (x^2 + 1) |u(x)|^2 \right) dx \\
& = & \sum_{n_1 = 0}^{\infty} \sum_{n_2 = 0}^{\infty} a_{n_1}
\bar{a}_{n_2} \int_{\mathbb{R}} \left[ \phi_{n_1}'(x)
\phi_{n_2}'(x) + ( x^2 + 1)\phi_{n_1}(x) \phi_{n_2}(x) \right] dx
\\ & = & 2 \sum_{n_1 = 0}^{\infty} \sum_{n_2 = 0}^{\infty}
a_{n_1} \bar{a}_{n_2} (1 + n_2) (\phi_{n_1},\phi_{n_2}) \\
& = & 2 \sum_{n = 0}^{\infty} (1 + n) |a_{n}|^2 = 2 \| {\bf a}
\|^2_{l^2_{1/2}},
\end{eqnarray*}
where the orthogonality relations (\ref{orthogonality}) have been
used.
\end{proof}

\begin{remark}
{\rm By the same method, one can prove that $u \in {\cal
H}_2(\mathbb{R})$ if and only if ${\bf a} \in l^2_1(\mathbb{N})$.}
\label{remark-embedding}
\end{remark}

\begin{lemma}
The vector field of the dynamical system (\ref{discrete-system})
maps $l^2_{1/2}(\mathbb{N})$ to $l^2_{-1/2}(\mathbb{N})$.
\label{lemma-phase-space}
\end{lemma}

\begin{proof}
The vector field of the dynamical system (\ref{discrete-system})
is decomposed into the linear ${\bf f}({\bf a})$ and nonlinear
$\sigma {\bf g}({\bf a})$ parts, where
$$
f_n = n a_n, \;\; g_n = \sum\limits_{(n_1,n_2,n_3)}
K_{n,n_1,n_2,n_3} a_{n_1} \bar{a}_{n_2} a_{n_3}, \quad \forall n =
0,1,2,3,...
$$
The linear unbounded part satisfies the estimate
\begin{equation}
\label{linear-vector-field} \| {\bf f}({\bf a}) \|^2_{l^2_s} =
\sum_{n = 0}^{\infty} ( 1 + n)^{2s} n^2 |a_n|^2 \leq \| {\bf a}
\|^2_{l^2_{s+1}},
\end{equation}
such that ${\bf f} : l^2_{s+1}(\mathbb{N}) \mapsto
l^2_s(\mathbb{N})$ for all $s \in \mathbb{R}$. If ${\bf a} \in
l^2_{1/2}(\mathbb{N})$, then $s = -\frac{1}{2}$. The nonlinear
vector part satisfies the estimate
\begin{eqnarray*}
\| {\bf g}({\bf a}) \|^2_{l^2_s} & = & \sum_{n = 0}^{\infty} ( 1 +
n)^{2s} \sum_{(n_1,n_2,n_3)} \sum_{(m_1,m_2,m_3)}
K_{n,n_1,n_2,n_3} K_{n,m_1,m_2,m_3} a_{n_1} \bar{a}_{n_2} a_{n_3}
\bar{a}_{m_1} a_{m_2} \bar{a}_{m_3} \\
& = & \sum_{n = 0}^{\infty} ( 1 + n)^{2s} \left| \left( \phi_n u,
|u|^2 \right) \right|^2 \leq \left( \sum_{n = 0}^{\infty} ( 1 +
n)^{2s} \| u \phi_n \|^2_{L^2} \right) \| u \|^4_{L^4} \\ & \leq &
\left( \sum_{n = 0}^{\infty} ( 1 + n)^{2s} \| \phi_n \|^2_{L^4}
\right) \| u \|^6_{L^4},
\end{eqnarray*}
where $u(x) = \sum\limits_{n = 0}^{\infty} a_n \phi_n(x)$ and all
$\phi_n(x)$ are real-valued. By the main theorem of \cite{Freud},
there exists a constant $C > 0$ such that
\begin{equation}
\label{estimate-2} \| \phi_n \|^4_{L^4} \leq C
\frac{\log(1+n)}{\sqrt{1 + n}}, \qquad \forall n = 0,1,2,...
\end{equation}
Therefore, the series $\sum_{n = 0}^{\infty} ( 1 + n)^{2s} \|
\phi_n \|^2_{L^4}$ converges for all $s < -\frac{3}{8}$. The value
$s = -\frac{1}{2}$ belongs to this interval. Finally, by the
Sobolev embedding and Poincare inequality \cite{Adams}, there are
constants $C, \tilde{C} > 0$ such that
$$
\| u \|^4_{L^4} \leq C \| (u^2)' \|^2_{L^2} \leq 4 C \| u
\|^2_{L^{\infty}} \| u' \|^2_{L^2} \leq \tilde{C} \| u \|^4_{H^1}
\leq \tilde{C} \| u \|^4_{{\cal H}_1}.
$$
Since the norm in ${\cal H}_1(\mathbb{R})$ for the function $u(x)$
is equivalent to the norm in $l^2_{1/2}(\mathbb{Z})$ for the
vector ${\bf a}$ by Lemma \ref{lemma-embedding}, the estimate for
the nonlinear vector field is completed by
\begin{equation}
\label{nonlinear-vector-field}  \| {\bf g}({\bf a})
\|^2_{l^2_{-1/2}} \leq C_0 \| u \|^6_{L^4} \leq \tilde{C}_0 \|
{\bf a} \|^6_{l^2_{1/2}},
\end{equation}
for some $C_0, \tilde{C}_0 > 0$. The interpolation argument for
the bounds (\ref{linear-vector-field}) and
(\ref{nonlinear-vector-field}) concludes the proof that the
nonlinear vector field ${\bf f}({\bf a}) + \sigma {\bf g}({\bf
a})$ maps $l^2_{1/2}(\mathbb{N})$ to $l^2_{-1/2}(\mathbb{N})$.
\end{proof}

\begin{theorem}
\label{theorem-equivalence} The discrete dynamical system
(\ref{discrete-system}) is globally well-posed in the phase space
${\bf a} \in l^2_{1/2}(\mathbb{N})$.
\end{theorem}

\begin{proof}
By Proposition 2.2 in \cite{Carles}, the GP equation
(\ref{GP-zero-epsilon}) is globally well-posed in the phase space
$u \in {\cal H}_1(\mathbb{R})$. By Lemma \ref{lemma-embedding},
the trajectory $u(t) \in {\cal H}_1(\mathbb{R})$ is equivalent to
the trajectory ${\bf a}(t) \in l^2_{1/2}(\mathbb{N})$ on $t \in
\mathbb{R}$. By Lemma \ref{lemma-phase-space}, the vector field of
the discrete dynamical system (\ref{discrete-system}) is
well-defined on $l^2_{1/2}(\mathbb{N}) \subset l^2(\mathbb{N})$,
where it is equivalent to the vector field of the GP equation
(\ref{GP-zero-epsilon}) by virtue of standard orthogonal
projections.
\end{proof}

\section{Existence and stability of stationary solutions}

Stationary solutions of the dynamical system
(\ref{discrete-system}) take the form ${\bf a}(t) = {\bf A} e^{-i
\mu t}$, where ${\bf A}$ is a time-independent vector and $\mu$ is
a parameter of the solution. If ${\bf A} \in
l^2_{1/2}(\mathbb{N})$ and $\phi(x) = \sum\limits_{n=0}^{\infty}
A_n \phi_n(x)$, then $\phi \in {\cal H}_1(\mathbb{R})$ is a
stationary solution of the GP equation (\ref{GP-zero-epsilon}),
that is $\phi(x)$ satisfies the ODE (\ref{ODE}). The vector ${\bf
A}$ is found as a root of the infinite-dimensional cubic vector
field ${\bf F} : l^2_{1/2}(\mathbb{N}) \times \mathbb{R} \mapsto
l^2_{-1/2}(\mathbb{N})$, where the $n$-th component of ${\bf
F}({\bf A},\mu)$ is given by
\begin{equation}
\label{stationary-solution} F_n = (\mu - n) A_n - \sigma
\sum_{(n_1,n_2,n_3)} K_{n;n_1,n_2,n_3} A_{n_1} \bar{A}_{n_2}
A_{n_3} = 0, \qquad \forall n = 0,1,2,...
\end{equation}
The Jacobian operator $D_{\bf A} {\bf F}({\bf 0},\mu)$ is a
diagonal matrix with entries $\mu - n$ and it admits a
one-dimensional kernel if $\mu = n_0$ for any non-negative integer
$n_0$. The corresponding eigenvector is ${\bf e}_{n_0}$, the unit
vector in $l^2(\mathbb{N})$. By the local bifurcation theory
\cite{GS}, each eigenvector of $D_{\bf A} {\bf F}({\bf 0},n_0)$
can be uniquely continued in a local neighborhood of the point
${\bf A} = {\bf 0} \in l^2_{1/2}(\mathbb{N})$ and $\mu = n_0 \in
\mathbb{R}$. We are particularly interested in the second
eigenvalue $n_0 = 1$, which corresponds to the {\em dark soliton}
$\phi(x)$ with a single zero (node) at $x = 0$. (Other
bifurcations of stationary localized solutions $\phi(x)$ are
considered in \cite{Alfimov}.) Details of this bifurcation are
given in the following proposition.

\begin{proposition}
\label{lemma-stationary} Consider real-valued roots $({\bf
A},\mu)$ of the vector field ${\bf F}({\bf A},\mu)$ such that
${\bf A} \in l^2_{1/2}(\mathbb{N})$. There exists a unique family
of solutions near $\mu = 1$ parameterized by $\varepsilon$ such
that
\begin{equation}
\label{comparison1} \| {\bf A} - \varepsilon {\bf e}_1
\|_{l^2_{1/2}} \leq C_1 \varepsilon^3, \qquad \left| \mu -  1 -
\frac{3 \sigma \varepsilon^2}{\sqrt{32 \pi}} \right| \leq C_2
\varepsilon^4,
\end{equation}
for some $\varepsilon$-independent constants $C_1, C_2 > 0$ and
sufficiently small $\varepsilon$. Moreover, if $\sigma \neq 0$,
the solution $({\bf A},\mu)$ is smooth with respect to
$\varepsilon$ for sufficiently small $\varepsilon$ and $\frac{d}{d
\mu} Q({\bf A}) \neq 0$, where $Q({\bf A}) = \| {\bf A}
\|^2_{l^2}$.
\end{proposition}

\begin{proof}
Both ${\bf F}({\bf A},\mu)$ and $D_{\bf A} {\bf F}({\bf A},\mu)$
are continuous in a local neighborhood of ${\bf A} = {\bf 0} \in
l^2_{1/2}(\mathbb{N})$ and $\mu = 1 \in \mathbb{R}$. At the point
${\bf A} = {\bf 0}$ and $\mu = 1$, the operator has a
one-dimensional kernel with the eigenvector ${\bf e}_1 \in
l^2(\mathbb{N})$. By using the method of Lyapunov--Schmidt
reductions \cite{GS}, we set ${\bf A} = \varepsilon \left[ {\bf
e}_1 + \tilde{\bf A} \right]$ and $\mu = 1 + \tilde{\mu}$, where
$\tilde{\bf A}$ is an orthogonal complement of ${\bf e}_1$ in
$l^2(\mathbb{N})$ such that $\tilde{A}_1 = 0$. The orthogonal
projection of equation (\ref{stationary-solution}) to ${\bf e}_1$
gives a bifurcation equation for $\tilde{\mu}$
\begin{eqnarray*}
\tilde{\mu} = \sigma \varepsilon^2 \left[ K_{1;1,1,1} + 3
\sum_{n_1} K_{1;1,1,n} \tilde{A}_{n_1} + 3 \sum_{(n_1,n_2)}
K_{1;1,n_1,n_2} \tilde{A}_{n_1} \tilde{A}_{n_2}+
\sum_{(n_1,n_2,n_3)} K_{1;n_1,n_2,n_3} \tilde{A}_{n_1}
\tilde{A}_{n_2} \tilde{A}_{n_3} \right],
\end{eqnarray*}
where the index for $(n_1,n_2,n_3)$ in the summation signs runs on
the set $\{ 0,2,3,... \}$. Let $P$ be an orthogonal projection
from $l^2(\mathbb{N})$ to the orthogonal complement of ${\bf
e}_1$. Then the inverse of $P D_{\bf A} {\bf F}({\bf 0},1)  P$
exists and is a bounded operator from $l^2_{1/2}(\mathbb{N})$ to
$l^2_{1/2}(\mathbb{N})$. By the Implicit Function Theorem, there
exists a unique smooth solution $\tilde{\bf A}$ in the
neighborhood of $\tilde{\bf A} = {\bf 0} \in l^2_s(\mathbb{N})$
such that $\| \tilde{\bf A} \|_{l^2_{1/2}} \leq C_1 \varepsilon^2$
for some $C_1 > 0$. By the Implicit Function Theorem, there exists
a unique smooth solution $\tilde{\mu}$ of the bifurcation equation
in the neighborhood of $\tilde{\mu} = 0$ such that $|\tilde{\mu} -
\varepsilon^2 \sigma K_{1,1,1,1} | \leq C_2 \varepsilon^4$ for
some $C_2 > 0$. The value $K_{1,1,1,1} = \| \phi_1\|^4_{L^4} =
\frac{3}{\sqrt{32 \pi}}$ is computed in Table I. Since $Q({\bf A})
= \| {\bf A} \|^2_{l^2} = \varepsilon^2 + {\rm O}(\varepsilon^4)$
and $\mu - 1 = \frac{3 \sigma \varepsilon^2}{\sqrt{32 \pi}} + {\rm
O}(\varepsilon^4)$, then $\frac{d}{d \mu} Q({\bf A}) \neq 0$ near
$\mu = 1$ for $\sigma \neq 0$.
\end{proof}

\begin{center}
\begin{tabular}{|c|l|l|l|l|l|l|}
\hline $\phantom{t}$ & $n = 0$ & $n = 1$ & $n = 2$ & $n = 3$ & $n = 4$ & $n = 5$ \\
\hline $K_{n,n,n,n}$  & $\frac{1}{\sqrt{2\pi}}$   & $\frac{3}{4
\sqrt{2\pi}}$ & $\frac{41}{64 \sqrt{2\pi}}$
& $\frac{147}{256 \sqrt{2\pi}}$ & $\frac{8649}{16384 \sqrt{2\pi}}$ & $\frac{32307}{65536 \sqrt{2\pi}}$\\
\hline $K_{1,n,n,1}$  & $\frac{1}{2 \sqrt{2\pi}}$   & $\frac{3}{4
\sqrt{2\pi}}$ & $\frac{7}{16 \sqrt{2\pi}}$
& $\frac{11}{32 \sqrt{2\pi}}$ & $\frac{75}{256 \sqrt{2\pi}}$ & $\frac{133}{512 \sqrt{2\pi}}$\\
\hline $K_{0,1,1,n}$ & $\frac{1}{2 \sqrt{2\pi}}$ & $0$ & $
\frac{1}{8 \sqrt{\pi}}$ & $0$ & $-\frac{3 \sqrt{3}}{32 \sqrt{\pi}}$ & $0$\\
\hline
\end{tabular}
\end{center}

{\bf Table I:} Numerical values for $K_{n,n,n,n} = \| \phi_n
\|^4_{L^4}$, $K_{1,n,n,1} = (\phi_1^2,\phi_n^2)$, and $K_{0,1,1,n}
= (\phi_0 \phi_n,\phi_1^2)$.

\noindent Let $({\bf A},\mu)$ be a real-valued root of the
nonlinear vector field (\ref{stationary-solution}) such that ${\bf
A} \in l^2_{1/2}(\mathbb{N})$. Spectral stability of the
stationary solution is studied with the expansion
\begin{equation}
\label{linear_stability} {\bf a}(t) = e^{-i \mu t} \left[ {\bf A}
+ \left( {\bf B} - {\bf C}\right) e^{i \Omega t} +  \left(
\bar{\bf B} + \bar{\bf C}\right) e^{- i \bar{\Omega} t} + {\rm
O}(\|{\bf B}\|^2 + \| {\bf C} \|^2) \right],
\end{equation}
where the spectral parameter $\Omega \in \mathbb{C}$ and the
eigenvector $({\bf B},{\bf C}) \in l^2(\mathbb{N},\mathbb{C}^2)$
satisfy the linear problem
\begin{equation}
\label{eigenvalue} L_+ {\bf B} = \Omega {\bf C}, \qquad L_- {\bf
C} = \Omega {\bf B},
\end{equation}
associated with matrix operators $L_{\pm}$. Their $n$-th
components are defined in the form
\begin{eqnarray}
\label{eigenvalue-1-2} \left\{ \begin{array}{ccc} (L_+ {\bf B})_n
& = & (n - \mu) B_n + 3 \sigma
\sum_{n_1} V_{n,n_1} B_{n_1}, \\
(L_- {\bf C})_n & = & (n - \mu ) C_n + \sigma \sum_{n_1} V_{n,n_1}
C_{n_1}, \end{array} \right. \quad \forall n = 0,1,2,3,...,
\end{eqnarray}
where $V_{n,n_1} = \sum\limits_{(n_2,n_3)} K_{n,n_1,n_2,n_3}
A_{n_2} A_{n_3}$. We have used here the symmetry of the
coefficients $K_{n,n_1,n_2,n_3}$ with respect to the interchange
of $(n_1,n_2,n_3)$.

\begin{lemma}
\label{lemma-operators} Let $({\bf A}, \mu)$ be a real-valued root
of the vector field ${\bf F}({\bf A},\mu)$ such that ${\bf A} \in
l^2_{1/2}(\mathbb{N})$. Operators $L_+$ and $L_-$ admit closed
self-adjoint extensions in $l^2(\mathbb{N})$ with the domain in
$l^2_1(\mathbb{N})$.
\end{lemma}

\begin{proof}
The diagonal unbounded part of $L_{\pm}$ maps $l^2_1(\mathbb{N})$
to $l^2(\mathbb{N})$. We need to show that the non-diagonal part
of $L_{\pm}$ represents a bounded perturbation from
$l^2(\mathbb{N})$ to $l^2(\mathbb{N})$ if ${\bf A} \in
l^2_{1/2}(\mathbb{N})$. This is done by using the same ideas as in
the proof of Lemma \ref{lemma-phase-space}:
\begin{eqnarray*}
\sum_{n=0}^{\infty} \left| \sum_{n_1} V_{n,n_1} B_{n_1} \right|^2
& = & \sum_{n=0}^{\infty} \sum_{(n_1,n_2,n_3)}
\sum_{(m_1,m_2,m_3)} K_{n,n_1,n_2,n_3} K_{n,m_1,m_2,m_3} A_{n_2}
A_{n_3} A_{m_2} A_{m_3} B_{n_1} \bar{B}_{m_1} \\
& = & \sum_{n = 0}^{\infty} |(\phi_n, u^2 v)|^2 = \| u^2 v
\|^2_{L^2} \leq \| u \|^4_{L^{\infty}} \| v \|^2_{L^2} \leq C^4 \|
u \|^4_{{\cal H}_1} \sum_{n = 0}^{\infty} |B_n|^2,
\end{eqnarray*}
where $u(x) = \sum_{n=0}^{\infty} A_n \phi_n(x)$ and $v(x) =
\sum_{n=0}^{\infty} B_n \phi_n(x)$.
\end{proof}

\begin{remark}
{\rm The result of Lemma \ref{lemma-operators} is obvious from the
equivalence between the space ${\cal H}_2(\mathbb{R})$ for the
function $v(x)$ and the space $l^2_1(\mathbb{N})$ for the vector
${\bf B}$, see Remark \ref{remark-embedding}. We recall that the
differential operators ${\cal L}_{\pm}$ given by (\ref{operators})
are defined on the domain ${\cal H}_2(\mathbb{R})$ and the matrix
operators $L_{\pm}$ given by (\ref{eigenvalue-1-2}) represent the
action of differential operators on the basis of Hermite functions
in ${\cal H}_2(\mathbb{R})$. }
\end{remark}

The linear problem (\ref{eigenvalue}) has eigenvalue $\Omega = 0$
of geometric multiplicity one and algebraic multiplicity two due
to the exact solution
\begin{equation}
\label{null-eigenvalue} L_- {\bf A} = {\bf 0}, \qquad L_+
\partial_{\mu} {\bf A} = {\bf A},
\end{equation}
where the smoothness of ${\bf A}$ with respect to $\mu$ near $\mu
= 1$ is guaranteed by Proposition \ref{lemma-stationary}.

When ${\bf A} = {\bf 0}$ and $\mu = 1$, the spectrum of the
eigenvalue problem (\ref{eigenvalue}) is known in the explicit
form. It consists of eigenvalues $\Omega = 0$ and $\Omega = \pm 1$
of geometric and algebraic multiplicities two and simple
eigenvalues $\Omega = \pm m$ for all $m = 2,3,...$. The double
zero eigenvalue persists for any $\varepsilon$ according to the
exact solution (\ref{null-eigenvalue}), stemming from the
underlying $U(1)$ invariance of the system. Splitting of all other
eigenvalues in a local neighborhood of ${\bf A} = {\bf 0}$ and
$\mu = 1$ is described by the following proposition.

\begin{proposition}
Let $({\bf A},\mu)$ be defined by Proposition
\ref{lemma-stationary} for sufficiently small $\varepsilon$.
Non-zero eigenvalues of the linear problem (\ref{eigenvalue}) form
a set $\{ \pm \Omega_m \}_{m = 0}^{\infty}$ of simple real
symmetric eigenvalue pairs, such that
\begin{eqnarray}
\label{eigenvalue-expansion-1}  \left| \Omega_0 - 1 \right| \leq
C_0 \varepsilon^4, \qquad \left| \Omega_1 - 1 + \frac{
\varepsilon^2 \sigma}{8 \sqrt{2 \pi}} \right| \leq C_1
\varepsilon^4
\end{eqnarray}
and
\begin{eqnarray}
\label{eigenvalue-expansion-3} \left| \Omega_m - m + \varepsilon^2
\sigma \left( K_{1,1,1,1} - 2 K_{m+1,1,1,m+1} \right) \right| \leq
C_m \varepsilon^4, \quad \forall m = 2,3,....
\end{eqnarray}
for some $\varepsilon$-independent constants $C_0,C_1,C_m > 0$.
\label{lemma-eigenvalue}
\end{proposition}

\begin{proof}
Since the essential spectrum of the matrix operators $L_{\pm}$ is
empty and the potential terms are bounded perturbations to the
unbounded diagonal terms, isolated eigenvalues split according to
the regular perturbation theory \cite{Kato}. The formal power
series expansion for a simple eigenvalue $\Omega = m = 2,3,...$ is
defined by
\begin{equation}
\left\{ \begin{array}{rcl} {\bf B} & = & {\bf e}_{m+1} +
\varepsilon^2 \tilde{\bf B} + {\rm O}(\varepsilon^4), \\ {\bf C} &
= & {\bf e}_{m+1} + \varepsilon^2 \tilde{\bf C} + {\rm
O}(\varepsilon^4), \\ \Omega & = & m + \varepsilon^2
\tilde{\Omega} + {\rm O}(\varepsilon^4).
\end{array} \right.
\end{equation}
Projections to the component $n = m+1$ lead to a linear system at
the leading order ${\rm O}(\varepsilon^2)$
\begin{eqnarray}
\left\{ \begin{array}{rcl} m \left( \tilde{B}_{m+1} -
\tilde{C}_{m+1} \right) & = & \sigma \left[ K_{1,1,1,1}
-  3 K_{m+1,1,1,m+1} \right] + \tilde{\Omega} \\
m \left( \tilde{C}_{m+1} - \tilde{B}_{m+1} \right) & = & \sigma
\left[ K_{1,1,1,1} - K_{m+1,1,1,m+1} \right] +\tilde{\Omega}.
\end{array} \right.
\end{eqnarray}
The linear system has a solution if and only if $\tilde{\Omega} =
\sigma \left( 2 K_{m+1,1,1,m+1} - K_{1,1,1,1}\right)$. Persistence
of the eigenvalue by the perturbation theory results in the
expansion (\ref{eigenvalue-expansion-3}). The power series
expansion for the double eigenvalue $\Omega = 1$ is defined by
\begin{eqnarray}
\left\{ \begin{array}{rcl} {\bf B} & = & \alpha {\bf e}_0 + \beta
{\bf e}_2 + \varepsilon^2 \tilde{\bf B} + {\rm O}(\varepsilon^4),
\\ {\bf C} & = & - \alpha {\bf e}_{0} + \beta {\bf e}_2 +
\varepsilon^2 \tilde{\bf C} + {\rm O}(\varepsilon^4), \\
\Omega & = & 1 + \varepsilon^2 \tilde{\Omega} + {\rm
O}(\varepsilon^4),
\end{array} \right.
\end{eqnarray}
where $(\alpha,\beta)$ are arbitrary parameters. Projections to
the components $n = 0$ and $n = 2$ leads to a linear system at the
leading order ${\rm O}(\varepsilon^2)$
\begin{eqnarray}
\left\{ \begin{array}{rcl} \left( \tilde{B}_{0} + \tilde{C}_{0}
\right) & = & \sigma \left[ 3 K_{0;1,1,0}
\alpha + 3 K_{0,1,1,2} \beta - K_{1,1,1,1} \alpha \right] + \tilde{\Omega} \alpha \\
- \left( \tilde{C}_{0} + \tilde{B}_0 \right) & = & \sigma \left[
K_{0,1,1,0} \alpha - K_{0,1,1,2} \beta - K_{1,1,1,1} \alpha
\right] + \tilde{\Omega} \alpha \\
\left( \tilde{B}_2 - \tilde{C}_2 \right) & = & \sigma \left[
K_{1,1,1,1} \beta - 3
K_{2,1,1,0} \alpha - 3 K_{2,1,1,2} \beta \right] + \tilde{\Omega} \beta \\
\left( \tilde{C}_2 - \tilde{B}_2 \right) & = & \sigma \left[
K_{1,1,1,1} \beta + K_{2,1,1,0} \alpha - K_{2,1,1,2} \beta \right]
+ \tilde{\Omega} \alpha \end{array} \right.
\end{eqnarray}
The linear system has a solution if and only if $(\alpha,\beta)$
satisfies a homogeneous system
\begin{eqnarray}
\label{hom-system-2} \left\{ \begin{array}{rcl} \sigma \left(
K_{1,1,1,1} \alpha - 2 K_{0,1,1,0} \alpha -
K_{0,1,1,2} \beta \right) & = & \tilde{\Omega} \alpha, \\
\sigma \left( -K_{1,1,1,1} \beta + K_{2,1,1,0} \alpha + 2
K_{2,1,1,2} \beta \right) & = & \tilde{\Omega} \beta.
\end{array} \right.
\end{eqnarray}
The homogeneous system for $(\alpha,\beta)$ has a non-zero
solution if and only if $\tilde{\Omega}$ satisfies a quadratic
equation, roots of which are given by
\begin{equation}
\tilde{\Omega} = \sigma \left( K_{2,1,1,2} - K_{0,1,1,0} \pm
\sqrt{(K_{1,1,1,1} - K_{0,1,1,0} - K_{2,1,1,2})^2 - K_{0,1,1,2}^2}
\right).
\end{equation}
It follows from Table I that $\sqrt{(K_{1,1,1,1} - K_{0,1,1,0} -
K_{2,1,1,2})^2 - K_{0,1,1,2}^2} = \frac{1}{16 \sqrt{2 \pi}}$ and
$K_{2,1,1,2} - K_{0,1,1,0} = -\frac{1}{16 \sqrt{2 \pi}}$.
Persistence of the eigenvalues by the perturbation theory results
in the expansion (\ref{eigenvalue-expansion-1}).
\end{proof}

\begin{corollary}
\label{corollary-signatures} Let $[{\bf B}_m,{\bf C}_m]^T$ be an
eigenvector of the linear problem (\ref{eigenvalue}) for the
eigenvalue $\Omega_m \in \mathbb{R}_+$ for any $m = 0,1,2,3...$ in
Proposition \ref{lemma-eigenvalue}. For sufficiently small
$\varepsilon$, the eigenvalue $\Omega_0$ has positive signature of
$\langle {\bf B}_0, L_+ {\bf B}_0 \rangle$, the eigenvalue
$\Omega_1$ has negative signature of $\langle {\bf B}_1, L_+ {\bf
B}_1 \rangle$, while all other eigenvalues $\Omega_m$ with $m =
2,3,...$ have positive signature of $\langle {\bf B}_m, L_+ {\bf
B}_m \rangle$, where $\langle \cdot, \cdot \rangle$ denotes a
standard inner product in $l^2(\mathbb{N})$.
\end{corollary}

\begin{proof}
In the case $\tilde{\Omega} = 0$, the homogeneous system
(\ref{hom-system-2}) for $(\alpha,\beta)$ has a one-parameter
family of solutions with $\beta = - \sqrt{2} \alpha$, such that
$\langle {\bf B}_0, L_+ {\bf B}_0 \rangle = - |\alpha|^2 +
|\beta|^2 + {\rm O}(\varepsilon^2) > 0$ for sufficiently small
$\varepsilon$. In the case $\tilde{\Omega} \neq 0$, the
homogeneous system (\ref{hom-system-2}) for $(\alpha,\beta)$ has a
one-parameter family of solutions with $\alpha = - \sqrt{2}
\beta$, such that $\langle {\bf B}_1, L_+ {\bf B}_1 \rangle = -
|\alpha|^2 + |\beta|^2 + {\rm O}(\varepsilon^2) < 0$ for
sufficiently small $\varepsilon$. In the case of other
eigenvalues, it is obvious from the proof of Proposition
\ref{lemma-eigenvalue} that $\langle {\bf B}_m, L_+ {\bf B}_m
\rangle = m + {\rm O}(\varepsilon^2)$ for $m = 2,3,...$.
\end{proof}

\begin{remark}
{\rm The double zero eigenvalue is associated with the expansions
\begin{equation}
\label{expansion-zero-eigenvalue} {\bf A} = \varepsilon {\bf e}_1
+ {\rm O}(\varepsilon^3), \qquad
\partial_{\mu} {\bf A} = \frac{\sqrt{32 \pi}}{3 \sigma \varepsilon}
{\bf e}_1 + {\rm O}(\varepsilon),
\end{equation}
where $\varepsilon$ is sufficiently small. As a result, $\langle
{\bf A}, \partial_{\mu} {\bf A} \rangle = \frac{\sqrt{32 \pi}}{3
\sigma} + {\rm O}(\varepsilon^2)$.} \label{remark-eigenvalues}
\end{remark}

\begin{lemma}
\label{lemma-diagonalization} Let $\varepsilon$ be sufficiently
small. Let $\{ [{\bf B}_m, {\bf C}_m]^T \}_{m = 0}^{\infty}$ be a
set of real-valued eigenvectors of the linear problem
(\ref{eigenvalue}) for the set of positive eigenvalues
$\{\Omega_m\}_{m=0}^{\infty}$. The set of eigenvectors is
symplectically orthogonal such that
\begin{equation}
\label{orthogonal-projections} \langle {\bf B}_{m'}, {\bf C}_m
\rangle = 0, \;\; \forall m' \neq m \qquad \langle {\bf B}_{m},
{\bf C}_m \rangle \neq 0, \;\; \forall m = 0,1,2,3...
\end{equation}
In addition, two eigenvectors $\{ [{\bf 0},{\bf A}]^T,
[\partial_{\mu} {\bf A},{\bf 0}]^T \}$ for the double zero
eigenvalue $\Omega = 0$ are symplectically orthogonal to other
eigenvectors and $\langle {\bf A},\partial_{\mu} {\bf A} \rangle
\neq 0$. The set of eigenvectors
\begin{equation}
\label{complete-set} \{ [{\bf B}_m,{\bf C}_m]^T \}_{m =
0}^{\infty} \oplus \{ [{\bf B}_m,-{\bf C}_m]^T \}_{m = 0}^{\infty}
\oplus \{ [{\bf 0},{\bf A}]^T, [\partial_{\mu} {\bf A},{\bf 0}]^T
\}
\end{equation}
is a basis in $l^2(\mathbb{N},\mathbb{R}^2)$ which is orthogonal
with respect to the symplectic projections
(\ref{orthogonal-projections}).
\end{lemma}

\begin{proof}
All eigenvalues $\{ \Omega_m \}_{m = 0}^{\infty}$ are positive and
simple for sufficiently small $\varepsilon$ by Proposition
\ref{lemma-eigenvalue}. Since $L_{\pm}$ are self-adjoint in
$l^2(\mathbb{N})$ and $\Omega_m$ is a real eigenvalue, then the
eigenvector $[{\bf B}_m,{\bf C}_m]^T$ of the linear problem
(\ref{eigenvalue}) can be chosen to be real-valued. The
orthogonality relations (\ref{orthogonal-projections}) follow by
direct computations from the linear problem (\ref{eigenvalue}) for
distinct eigenvalues $\Omega_{m'} \neq \Omega_m$ for all $m' \neq
m$. Values of $\langle {\bf B}_{m}, {\bf C}_m \rangle$ are
proportional to the values of $\langle {\bf B}_{m}, L_+ {\bf B}_m
\rangle$ for $\Omega_m \neq 0$ and they are non-zero for
sufficiently small $\varepsilon$ by Corollary
\ref{corollary-signatures}. The value of $\langle {\bf
A},\partial_{\mu} {\bf A} \rangle  = \frac{1}{2} \frac{d}{d \mu}
Q({\bf A})$ is non-zero for sufficiently small $\varepsilon$ by
Proposition \ref{lemma-stationary}. By the proof of Proposition
\ref{lemma-eigenvalue} and Remark \ref{remark-eigenvalues}, the
eigenvectors of the set (\ref{complete-set}) are represented for
sufficiently small $\varepsilon$ by the standard basis $\{ {\bf
e}_m \}_{m = 0}^{\infty} \oplus \{ {\bf e}_m \}_{m = 0}^{\infty}$
perturbed by a bounded perturbation in $l^2(\mathbb{N})$ of the
order ${\rm O}(\varepsilon^2)$. Also
\begin{equation}
\label{asymptotic-distribution-eigenvalues} \Omega_m = m + {\rm
O}(\varepsilon^2), \quad \langle {\bf B}_m, {\bf C}_m \rangle =
\frac{\langle {\bf B}_{m}, L_+ {\bf B}_m \rangle}{\Omega_m} = 1 +
{\rm O}(m^{-1} \varepsilon^2), \quad \forall m = 2,3,...,
\end{equation}
for sufficiently small $\varepsilon$, uniformly in $m$. Since no
other eigenvalues exist, the set of linearly independent
eigenvectors (\ref{complete-set}) is complete in
$l^2(\mathbb{N},\mathbb{R}^2)$. According to the Banach Theorem
for non-self-adjoint operators, the set is a basis if and only if
the spectral projections are bounded from below by a non-zero
constant in the limit $m \to \infty$, which follows from the
uniform asymptotic distribution
(\ref{asymptotic-distribution-eigenvalues}). Therefore, the set
(\ref{complete-set}) is a basis in $l^2(\mathbb{N},\mathbb{R}^2)$.
\end{proof}

\begin{lemma}
\label{lemma-persistence} Fix $\varepsilon \neq 0$ sufficiently
small. Simple positive eigenvalues of the set $\{\Omega_m \}_{m =
0}^{\infty}$ lie in the intervals
\begin{equation}
\label{distribution-1} \sigma > 0: \;\; \Omega_0 = 1 \;\;
\mbox{and} \;\; m - C_m^- \varepsilon^2 < \Omega_m < m, \;\;
\forall m \in \mathbb{N}
\end{equation}
and
\begin{equation}
\label{distribution-2} \sigma < 0: \;\; \Omega_0 = 1 \;\;
\mbox{and} \;\; m < \Omega_m < m + C_m^+ \varepsilon^2, \;\;
\forall m \in \mathbb{N}
\end{equation}
for some $\varepsilon$-independent constants $C_m^{\pm} > 0$.
\end{lemma}

\begin{proof}
The eigenvalue $\Omega_0 = 1$ persists for any $\varepsilon \in
\mathbb{R}$ due to equivalence of the linear eigenvalue problems
(\ref{spectrum}) and (\ref{eigenvalue}) for $\phi \in {\cal
H}_1(\mathbb{R})$ and ${\bf A} \in l^2_{1/2}(\mathbb{N})$ and the
existence of the exact solution (\ref{exact-solution-spectrum}) of
the linear eigenvalue problem (\ref{spectrum}). The corresponding
eigenvector $({\bf B},{\bf C})$ of the linear eigenvalue problem
(\ref{eigenvalue}) is found from the series representation
$$
\phi'(x) = \sum\limits_{n=0}^{\infty} B_n \phi_n(x), \qquad -x
\phi(x) = \sum\limits_{n=0}^{\infty} C_n \phi_n(x).
$$
The eigenvalue $\Omega_1$ satisfies the bounds
(\ref{distribution-1})--(\ref{distribution-2}) due to the explicit
bound (\ref{eigenvalue-expansion-1}). We use the bound
(\ref{eigenvalue-expansion-3}) to prove the bounds
(\ref{distribution-1})--(\ref{distribution-2}) for eigenvalues
$\Omega_m$ for all $m = 2,3,...$. The values of $K_{1,1,1,1} - 2
K_{m+1,1,1,m+1}$ are positive for the first values of $m =
3,4,...$ as follows from Table I, e.g.
$$
K_{1,1,1,1} - 2 K_{3,1,1,3} = \frac{1}{16 \sqrt{2\pi}}, \;\;
K_{1,1,1,1} - 2 K_{4,1,1,4} = \frac{21}{128 \sqrt{2\pi}}, \;\;
K_{1,1,1,1} - 2 K_{5,1,1,5} = \frac{59}{256 \sqrt{2 \pi}}.
$$
Note that the positive numerical values are monotonically
increasing. According to the main theorem in \cite{Freud}, the
sequence $\{ \| \phi_n\|_{L^4}\}_{n \in \mathbb{N}}$ is
monotonically decreasing to zero with the bound
(\ref{estimate-2}). Since $K_{m+1,1,1,m+1} \leq \| \phi_1
\|^2_{L^4} \| \phi_{m+1} \|^2_{L^4}$, then
$$
K_{1,1,1,1} - 2 K_{m+1,1,1,m+1} \geq \| \phi_1 \|^2_{L^4} \left(
\| \phi_1 \|^2_{L^4} - 2 \| \phi_{m+1} \|^2_{L^4} \right), \qquad
\forall m = 2,3,...
$$
Since $\| \phi_{m+1} \|^2_{L^4}$ decays monotonically to zero as
$m \to \infty$, there exists $M$ sufficiently large, such that the
lower bound above is strictly positive for $m \geq M$.
\end{proof}

\section{Existence of periodic solutions}

Let $({\bf A},\mu)$ be a real-valued root of the nonlinear vector
field ${\bf F}({\bf A},\mu)$ such that ${\bf A} \in
l^2_{1/2}(\mathbb{N})$. We use a decomposition ${\bf a}(t) = e^{-
i \mu t} \left[ {\bf A} + {\bf B}(t) + i {\bf C}(t) \right]$ with
real-valued vectors ${\bf B}$ and ${\bf C}$ to rewrite the
discrete dynamical system (\ref{discrete-system}) in the form
\begin{eqnarray}
\label{nonlinear-system} \dot{\bf B} = L_- {\bf C} + \sigma {\bf
N}_-({\bf B},{\bf C}), \qquad - \dot{\bf C} = L_+ {\bf B} + \sigma
{\bf N}_+({\bf B},{\bf C}),
\end{eqnarray}
where the operators $L_{\pm}$ are defined by
(\ref{eigenvalue-1-2}) and the vector fields ${\bf N}_{\pm}({\bf
B},{\bf C})$ contains quadratic and cubic terms with respect to
$({\bf B},{\bf C})$. By Theorem \ref{theorem-equivalence}, the
initial-value problem for system (\ref{nonlinear-system}) is
globally well-posed and the solution set $({\bf B},{\bf C}) \in
l^1_{1/2}(\mathbb{N},\mathbb{R}^2)$ is equivalent to the solution
set $(v,w) \in {\cal H}_1(\mathbb{R},\mathbb{R}^2)$ of the PDE
system (\ref{PDE}). The discrete dynamical system
(\ref{nonlinear-system}) inherits the Hamiltonian function
(\ref{Ham-function}) in the form
\begin{eqnarray}
\nonumber H & = & \frac{1}{2} \langle {\bf B},L_+ {\bf B} \rangle
+ \frac{1}{2} \langle {\bf C},L_- {\bf C} \rangle + \sigma
\sum_{(n,n_1,n_2,n_3)} K_{n,n_1,n_2,n_3} A_{n_1} (B_{n_2} B_{n_3}
+ C_{n_2} C_{n_3}) B_n
\\ \label{Ham-function-new} & + &  \frac{\sigma}{4} \sum_{(n,n_1,n_2,n_3)}
K_{n,n_1,n_2,n_3} \left( B_{n_1} B_{n_2} B_{n_3} B_n + 2 B_{n_1}
B_{n_2} C_{n_3} C_{n} + C_{n_1} C_{n_2} C_{n_3} C_n \right)
\end{eqnarray}
and the conserved quantity (\ref{gauge-fun}) in the form
\begin{equation}
\label{gauge-fun-new} Q = 2 \langle {\bf A},{\bf B} \rangle  +
\langle {\bf B},{\bf B} \rangle  + \langle {\bf C},{\bf C}
\rangle.
\end{equation}
Using Lemma \ref{lemma-diagonalization}, we represent a solution
$({\bf B},{\bf C})$ of the discrete system
(\ref{nonlinear-system}) by the series of eigenvectors
(\ref{complete-set}) associated with the linear problem
(\ref{eigenvalue}):
\begin{eqnarray}
\label{diagonalization-representaion} \left\{ \begin{array}{ccc}
{\bf B}(t) & = & \sum_{m = 0}^{\infty} b_m(t) {\bf B}_m +  \sum_{m
= 0}^{\infty} \bar{b}_m(t) {\bf B}_m  + \beta(t) \partial_{\mu} {\bf A}, \\
{\bf C}(t) & = & i \sum_{m = 0}^{\infty} b_m(t) {\bf C}_m - i
\sum_{m = 0}^{\infty} \bar{b}_m(t) {\bf C}_m  + \gamma(t) {\bf A},
\end{array} \right.
\end{eqnarray}
where $b_0(t)$, ${\bf b}(t) = (b_1,b_2,...)$ are complex-valued
and $\beta(t)$, $\gamma(t)$ are real-valued. The linear part of
system (\ref{nonlinear-system}) becomes block-diagonal in the
representation (\ref{diagonalization-representaion}), yielding the
evolution equations
\begin{equation}
\label{non-system-1} \dot{b}_m - i \Omega_m b_m = \sigma
N_m(b_0,{\bf b},\beta,\gamma), \quad \forall m = 0,1,2,3...
\end{equation}
and
\begin{equation}
\label{non-system-2} \dot{\beta} = \sigma S_0(b_0,{\bf
b},\beta,\gamma), \quad \dot{\gamma} + \beta = \sigma S_1(b_0,{\bf
b},\beta,\gamma),
\end{equation}
where
\begin{eqnarray*}
N_m(b_0,{\bf b},\beta,\gamma) = \frac{\langle {\bf C}_m, {\bf
N}_-({\bf B},{\bf C}) \rangle + i \langle {\bf B}_m, {\bf
N}_+({\bf B},{\bf C}) \rangle}{2 \langle {\bf C}_m,{\bf B}_m
\rangle}, \;\; \forall m = 0,1,2,3,...
\end{eqnarray*}
and
\begin{eqnarray*}
S_0(b_0,{\bf b},\beta,\gamma) = \frac{\langle {\bf A}, {\bf
N}_-({\bf B},{\bf C}) \rangle}{\langle {\bf A}, \partial_{\mu}
{\bf A}\rangle}, \qquad  S_1(b_0,{\bf b},\beta,\gamma) = -
\frac{\langle
\partial_{\mu} {\bf A}, {\bf N}_+({\bf B},{\bf C}) \rangle}{
\langle {\bf A}, \partial_{\mu} {\bf A}\rangle}.
\end{eqnarray*}
Using conservation of $Q$ given by (\ref{gauge-fun-new}) and the
decomposition (\ref{diagonalization-representaion}), one can
integrate the first equation of system (\ref{non-system-2}) in the
form
\begin{equation}
\label{non-system-3} \beta = \frac{Q - \| {\bf B} \|^2_{l^2} - \|
{\bf C} \|^2_{l^2}}{2 \langle {\bf A}, \partial_{\mu} {\bf
A}\rangle},
\end{equation}
where $Q$ is constant in time $t \in \mathbb{R}$. As a result, the
second equation of system (\ref{non-system-2}) is rewritten
explicitly in the form
\begin{equation}
\label{non-system-4} \dot{\gamma} = \frac{\| {\bf B} \|^2_{l^2} +
\| {\bf C} \|^2_{l^2} - 2 \sigma \langle
\partial_{\mu} {\bf A}, {\bf N}_+({\bf B},{\bf C}) \rangle - Q}{2 \langle {\bf A},
\partial_{\mu} {\bf A}\rangle}.
\end{equation}
We are now ready to apply the method of Lyapunov--Schmidt
reductions to the proof of Theorem \ref{theorem-main}.

\begin{proof1}{\em of Theorem \ref{theorem-main}}:
The vector space $({\bf B},{\bf C}) \in
l^2_{1/2}(\mathbb{N},\mathbb{R}^2)$ is equivalent to the vector
space ${\bf b} \in l^2_{1/2}(\mathbb{N})$ because of the
asymptotic distribution
(\ref{asymptotic-distribution-eigenvalues}). For instance, one
obtains that
$$
\sum_{n = 0}^{\infty} (1+n) |B_n| \sim \langle {\bf B}, L_+ {\bf
B} \rangle = 2 \sum_{m=0}^{\infty} \Omega_m \langle {\bf C}_m,
{\bf B}_m \rangle |b_m|^2 + |\beta|^2 \langle {\bf
A},\partial_{\mu} {\bf A} \rangle \sim \sum_{n \in \mathbb{N}}
(1+n) |b_n|^2.
$$
We should work in the space of $T$-periodic functions $b_0(t)$,
${\bf b}(t) \in l^2_{1/2}(\mathbb{N})$, $\beta(t)$ and $\gamma(t)$
on $t \in \mathbb{R}$, where $T$ is close to $2 \pi$. This period
corresponds to the eigenvalue $\Omega_0 = 1$ which persists for
any $\varepsilon \in \mathbb{R}$. By Lemma
\ref{lemma-persistence}, all other eigenvalues of the linear
problem (\ref{eigenvalue}) satisfy the non-resonance conditions $n
\neq \Omega_m$, $\forall n,m \in \mathbb{N}$ for any fixed
$\varepsilon \neq 0$ sufficiently small. As a result, we define
periodic functions ${\bf b}(t)$, $\beta(t)$ and $\gamma(t)$ in
terms of the periodic function $b_0(t)$, which solves a reduced
evolution problem. Let $\delta$ be sufficiently small. We shall
prove that there exist solutions of system (\ref{non-system-1}),
(\ref{non-system-3}) and (\ref{non-system-4}) which are
$T$-periodic on $t \in \mathbb{R}$ satisfying the apriori bounds
\begin{equation}
\label{apriori-bound} |b_0(t)| \leq \varepsilon \delta C_0, \;\;
\| {\bf b}(t) \|_{l^2_{1/2}} \leq \varepsilon \delta^2 C_b, \;\;
|\beta(t)| \leq \varepsilon^2 \delta^2 C_{\beta}, \;\; |\gamma(t)
- \delta \alpha| \leq \varepsilon^2 \delta^2 C_{\gamma}, \;\;
\forall t \in \mathbb{R}, \; \forall \alpha \in \mathbb{R},
\end{equation}
for some ($\varepsilon,\delta$)-independent constants
$C_0,C_b,C_{\beta},C_{\gamma} > 0$. If $b_0(t)$, ${\bf b}(t) \in
l^2_{1/2}(\mathbb{N})$, $\beta(t)$ and $\gamma(t)$ are
$T$-periodic functions on $t \in \mathbb{R}$ satisfying the bounds
(\ref{apriori-bound}), then $({\bf B}(t),{\bf C}(t)) \in
l^2_{1/2}(\mathbb{N},\mathbb{R}^2)$ is a $T$-periodic function on
$t \in \mathbb{R}$ satisfying the bound
\begin{equation}
\| {\bf B}(t)\|_{l^2_{1/2}} + \| {\bf C}(t) \|_{l^2_{1/2}} \leq C
\varepsilon \delta, \qquad \forall t \in \mathbb{R}, \; \forall
\alpha \in \mathbb{R},
\end{equation}
for some ($\varepsilon,\delta$)-independent constant $C > 0$. Here
we recall the expansion (\ref{expansion-zero-eigenvalue}) for
${\bf A}$, $\partial_{\mu} {\bf A}$ and the fact that $({\bf
B}_m,{\bf C}_m)^T$ are close to the unit vectors ${\bf e}_m$ for
sufficiently small $\varepsilon$. Since ${\bf N}_{\pm}({\bf
B},{\bf C})$ is cubic with respect $({\bf A},{\bf B},{\bf C})$,
contains quadratic terms in $({\bf B},{\bf C})$, and maps
$l^2_{1/2}(\mathbb{N},\mathbb{R}^2)$ to
$l^2_{-1/2}(\mathbb{N},\mathbb{R}^2)$, we obtain the bound
\begin{equation}
\label{bound-N-plus} \| {\bf N}_{\pm}({\bf B}(t),{\bf C}(t))
\|_{l^2_{-1/2}} \leq C_{\pm} \varepsilon^3 \delta^2, \qquad
\forall t \in \mathbb{R}, \; \forall \alpha \in \mathbb{R},
\end{equation}
for some ($\varepsilon,\delta$)-independent constants $C_{\pm} >
0$. By the Implicit Function Theorem to the right-hand-side of
equation (\ref{non-system-4}), there exists a unique constant $Q$
in the interval $|Q| \leq C_Q \varepsilon^2 \delta^2$ for some
$C_Q > 0$, such that the periodic function in the right-hand-side
of equation (\ref{non-system-4}) has zero mean on $t \in
\mathbb{R}$. In this case, there exists a periodic solution
$\gamma(t) = \delta \alpha + \tilde{\gamma}(t)$ of the
differential equation (\ref{non-system-4}), where
$\tilde{\gamma}(t)$ is a uniquely defined varying part and $\delta
\alpha$ is an arbitrary mean part. The varying part
$\tilde{\gamma}(t)$ satisfies the last bound in the list
(\ref{apriori-bound}). The function $\beta(t)$ is uniquely defined
by the explicit representation (\ref{non-system-3}) and it hence
satisfies the third bound in the list (\ref{apriori-bound}).

\noindent Consider now system (\ref{non-system-1}) for $m \in
\mathbb{N}$. Recall that $\Omega_m - m = {\rm O}(\epsilon^2)$ for
$m = 1,2,...$ uniformly in $m \in \mathbb{N}$ for sufficiently
small $\varepsilon$. By the Implicit Function Theorem, there
exists a unique solution ${\bf b}(t) \in l^2_{1/2}(\mathbb{N})$
defined by the periodic function $b_0(t)$ and parameter $\alpha
\in \mathbb{R}$ for sufficiently small $\delta$ provided that the
distance $|\Omega_m - m| \neq 0$ and the frequency $\Omega$ of the
periodic function $b_0(t)$ is such that $\Omega \to 1$ as $\delta
\to 0$. By the bound (\ref{bound-N-plus}) and the distribution
$\Omega_m - m = {\rm O}(\epsilon^2)$ for all $m \in \mathbb{N}$,
the function ${\bf b}(t)$ satisfies the second bound in the list
(\ref{apriori-bound}).

\noindent Eliminating the components ${\bf b}$, $\beta$ and
$\gamma$ from equation (\ref{non-system-1}) for $n = 0$, we obtain
a reduced evolution problem for $b_0(t)$ in the form
\begin{equation}
\label{normal-form} \dot{b}_0 = i b_0 + R(b_0;\alpha),
\end{equation}
where $R(b_0;\alpha)$ is a remainder term. Explicit computations
of $N_0(b_0,{\bf b},\beta,\gamma)$ show that
\begin{eqnarray}
\nonumber R(b_0;\alpha) & = & \varepsilon \left[ i
K_1(\varepsilon) b_0^2 + i K_2(\varepsilon) \bar{b}_0^2 + i
K_3(\varepsilon) |b_0|^2 + i K_4(\varepsilon) \delta^2 \alpha^2 +
K_5(\varepsilon) \delta \alpha \bar{b}_0 \right] \\
\label{remainder-term} & \phantom{t} & \phantom{text} + {\rm
O}\left(|b_0|^3,\varepsilon^2 \delta^2 \alpha^2 |b_0|,\varepsilon
|b_0| \| {\bf b}\| \right),
\end{eqnarray}
where $K_{1,2,3,4,5}$ are real-valued constants which are bounded
for sufficiently small $\varepsilon$. We are looking for
$T$-periodic functions $b_0(t)$ which satisfy the evolution
problem (\ref{normal-form}), have the leading order $b_0 \sim
\varepsilon \delta e^{i t + i \tau}$, where $\tau \in \mathbb{R}$
is arbitrary, and satisfy the first bound in the list
(\ref{apriori-bound}). By the normal form analysis of the ODE
(\ref{normal-form}) (see \cite{K95}), the quadratic terms in the
remainder (\ref{remainder-term}) do not change the frequency
$\Omega$ of oscillations of the periodic function $b_0(t)$ at the
leading order and therefore, $|\Omega - 1| \leq C_{\Omega}
\varepsilon^2 \delta^2$ for some $C_{\Omega} > 0$. Since the
Hamiltonian function (\ref{Ham-function-new}) of system
(\ref{nonlinear-system}) is constant in time, it remains constant
when the function $b_0(t)$ solves the reduced evolution problem
(\ref{normal-form}) and the functions ${\bf b}(t)$, $\beta(t)$ and
$\gamma(t)$ are constructed above. By the normal form analysis of
reversible systems, there exists a two-dimensional invariant
manifold of system (\ref{normal-form}) filled with periodic
solutions of frequencies close to $\Omega = 1$ and parameterized
by $(\delta,\tau)$ in addition to parameter
$(\varepsilon,\alpha)$.
\end{proof1}

\begin{remark}
{\rm Theorem \ref{theorem-main} is reminiscent of an
infinite-dimensional analogue of the Lyapunov Theorem for
persistence of periodic orbits in Hamiltonian systems (see Chapter
II, Section 45 on pp. 166--180 of \cite{Lyap}). However, due to
the symmetries, a double zero eigenvalue occurs in the linear
problem (\ref{eigenvalue}), and the proof of Theorem
\ref{theorem-main} is complicated by the analysis of the
associated two-dimensional subspace. Similar theorems on
persistence of $k$-dimensional tori in $n$-dimensional Hamiltonian
system with $k-1$ additional conserved quantities were studied in
the Nekhoroshev--Kuksin Theorems (see Theorem 2.3 on p. 4 of
\cite{BG} and Theorem 1 on p. xiii of \cite{Kuksin}).}
\end{remark}

\begin{remark}
{\rm The periodic solution of Theorem \ref{theorem-main} has the
smallest frequency in the focusing case $\sigma = -1$, since
$\Omega_1 > 1$ in the bound (\ref{distribution-2}) for
sufficiently small $\varepsilon$. However, it is not the smallest
frequency in the defocusing case $\sigma = 1$ since $\Omega_1 < 1$
in the bound (\ref{distribution-1}). Persistence of the periodic
solution for the smallest frequency $\Omega_1$ can not be proved
by a simple application of the Lyapunov Theorem since the bound
(\ref{distribution-1}) does not guarantee that the non-resonance
conditions $n \Omega_1 \neq \Omega_m$ are satisfied for all $n \in
\mathbb{N}$ and $m = 2,3,...$. By the same reason, persistence of
quasi-periodic oscillations on the tori with two and more
frequencies $\{ 1,\Omega_1,\Omega_2,... \}$ can not be proved for
small $\varepsilon$. }
\end{remark}

\begin{remark}
{\rm Persistence of quasi-periodic oscillations on the tori along
the Cantor set of parameter values was proved in Section 2.5 on p.
33 of \cite{Kuksin} for the Hartree nonlinear functions and a
perturbation of the parabolic potential $V(x) = \frac{1}{2} x^2$
by a localized potential $V_0(x)$. Our main result is stronger
than this application of the main theorem in \cite{Kuksin} since
the periodic orbit is continuous with respect to parameters of the
PDE problem rather than along the Cantor set of parameter values.}
\end{remark}

\section{Numerical Results}

\begin{figure}
\begin{center}
\includegraphics[height=8cm]{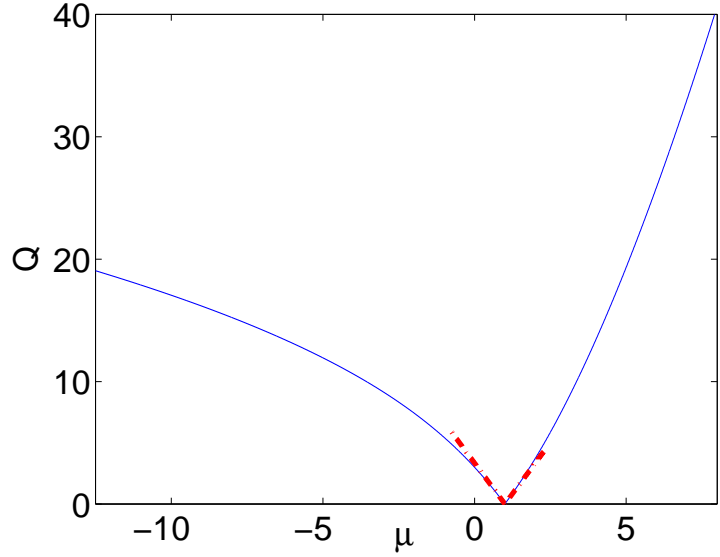}
\end{center}
\caption{Branches of dark solitons versus $\mu$ both for the case
of $\sigma=-1$ (when $\mu<1$) and $\sigma=1$ (when $\mu>1$). The
numerically obtained solution is shown by solid line and the
asymptotic solution (\ref{comparison1}) is shown by dash-dotted
line.} \label{dfig1}
\end{figure}

We illustrate results of our manuscript with some numerical
approximations. First, we identify the relevant branch of
stationary solutions of the ODE (\ref{ODE}). To do so, we use a
fixed point method (Newton-Raphson iteration) to solve a
discretized boundary-value problem. A centered-difference scheme
is applied to the second-order derivatives with a typical spacing
$\Delta x \in [0.025,0.1]$. We are using a sufficiently large
computational domain $x \in [-L,L]$ such that the boundary
conditions do not affect the approximations within the considered
numerical precision. The solutions $\phi(x)$ are obtained, using
continuation, as a function of parameter $\mu$. The continuation
of the solution branches is performed from the linear limit $\mu =
1$, both for the cases $\sigma=1$ and $\sigma=-1$. The results are
shown in Figure \ref{dfig1}, illustrating the quantity $Q = \|
\phi \|^2_{L^2}$ as a function of $\mu$. The numerical findings
are also compared to the asymptotic result (\ref{comparison1}) of
Proposition \ref{lemma-stationary} indicating the good agreement
of the latter prediction with our computational results for a
fairly wide parametric window.

\begin{figure}
\begin{center}
\includegraphics[height=8cm]{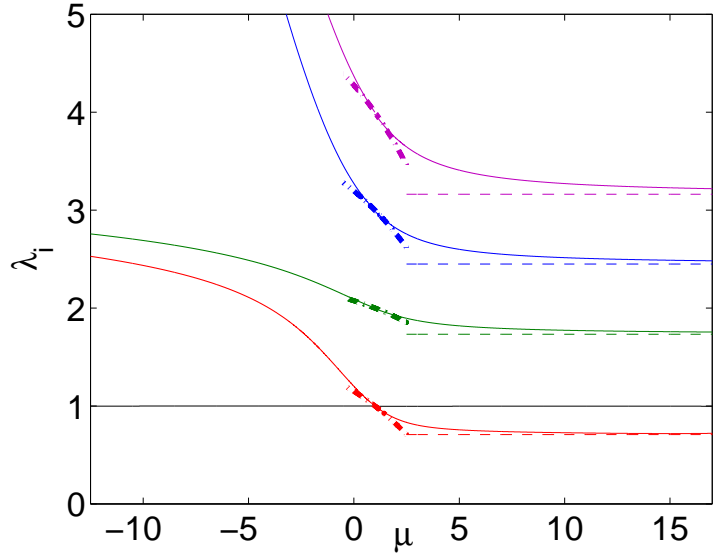}
\end{center}
\caption{Smallest purely imaginary eigenvalues of the linear
eigenvalue problem (\ref{spectrum}) versus $\mu$. The numerically
obtained eigenvalues are shown by solid lines, the asymptotic
results
(\ref{eigenvalue-expansion-1})-(\ref{eigenvalue-expansion-3}) are
shown by dash-dotted lines, and the asymptotic results
(\ref{eigenvalue-expansion-1-asymptotic})--(\ref{eigenvalue-expansion-3-asymptotic})
are shown by dashed lines. } \label{dfig2}
\end{figure}

Once the corresponding numerical solution is identified (for a
given $\sigma$ and $\mu$), the linear eigenvalue problem
(\ref{spectrum}) is approximated numerically. We use again a
discretization of differential operators on a finite grid, such
that the spectral problem (\ref{spectrum}) becomes a matrix
eigenvalue problem that is solved through standard numerical
linear algebra routines. The relevant lowest eigenvalues are
presented in Figure \ref{dfig2} and are also compared with the
corresponding asymptotic results
(\ref{eigenvalue-expansion-1})--(\ref{eigenvalue-expansion-3}) of
Proposition \ref{lemma-eigenvalue}. The dashed lines show
asymptotic results
(\ref{eigenvalue-expansion-1-asymptotic})--(\ref{eigenvalue-expansion-3-asymptotic})
of Appendix A derived in the limit $\mu \to \infty$ for $\sigma =
1$. Once again, the good agreement offers us a quantitative handle
on the relevant eigenvalues.

\begin{figure}
\begin{center}
\includegraphics[height=14cm]{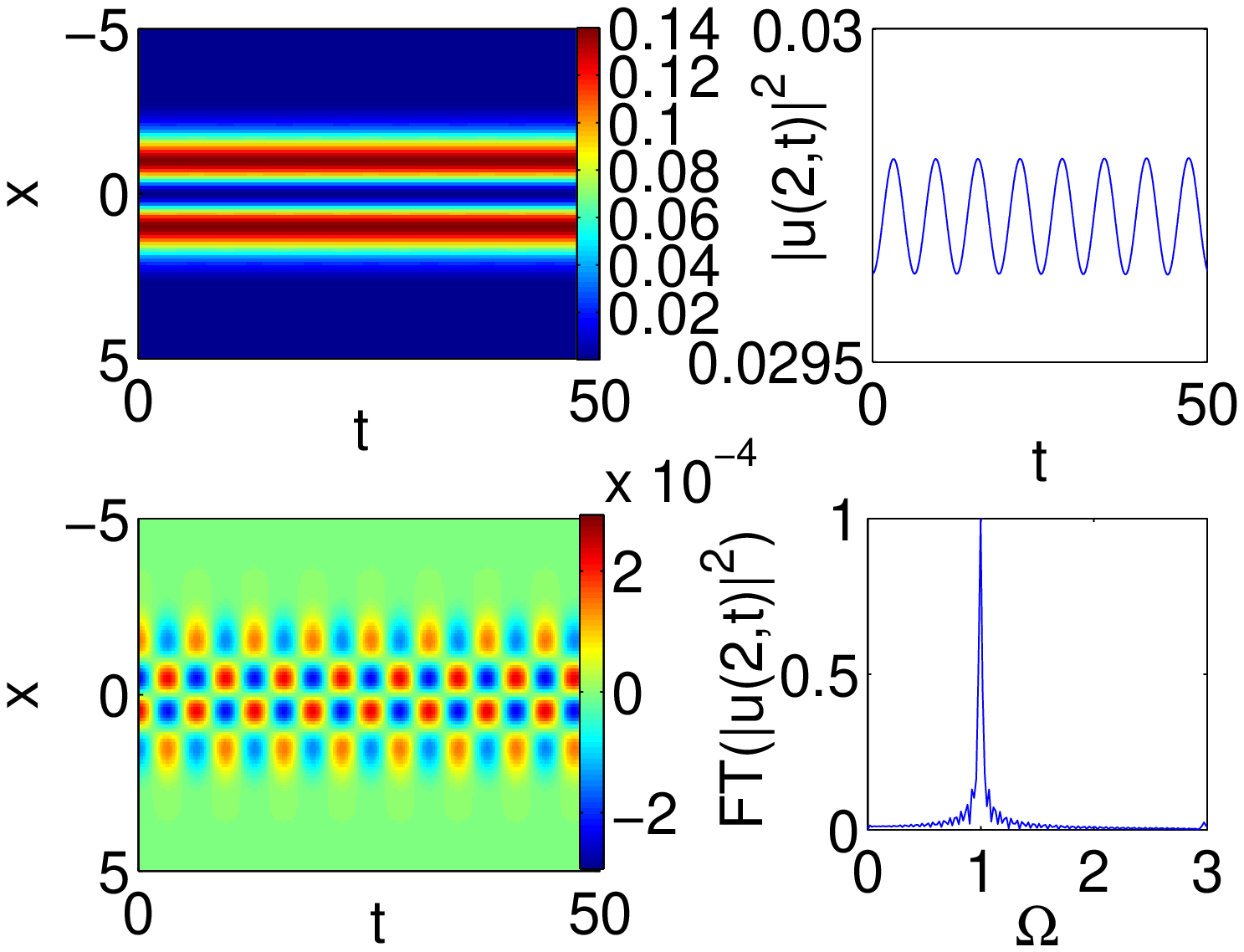}
\end{center}
\caption{A typical example of the robust time-periodic solution of
the Gross-Pitaevskii equation (\ref{GP-zero-epsilon}) for $\sigma
= 1$, $\mu = 1.1$ and $u(x,0) = \phi(x) + \delta \phi'(x)$ with
$\delta = 10^{-3}$. The top left panel shows the space-time
contour plot of $|u(x,t)|^2$, the bottom left panel shows the
space-time contour of $|u(x,t)|^2-\phi^2(x)$. The top right panel
shows the time evolution of $|u(x_0,t)|^2$ with $x_0 = 2$, while
the bottom right panel shows the Fourier transform of the time
series of $|u(x_0,t)|^2$, featuring a peak at $\Omega \approx 1$.}
\label{dfig3}
\end{figure}

Finally, we have also examined periodic oscillations of dark
solitons in the numerical simulations of the GP equation
(\ref{GP-zero-epsilon}). A typical example is shown in Figure
\ref{dfig3} for $\sigma = 1$ and $\mu = 1.1$ for the initial
condition $u(x,0) = \phi(x) + \delta \phi'(x)$ with $\delta =
10^{-3}$. The top left panel shows the space-time contour plot of
$|u(x,t)|^2$, clearly highlighting that this is a small
(imperceptible, at the scale of this panel) perturbation of a
stable stationary solution $\phi(x)$. The bottom left panel shows
the space-time contour plot of $|u(x,t)|^2 - \phi^2(x)$,
emphasizing the time-periodic oscillations of the perturbation to
the stationary solution. The periodic oscillations are also
visible on the top right panel where $|u(x_0,t)|^2$ is plotted
versus $t$ for $x_0 = 2$. Finally, the bottom right panel
illustrates the Fourier transform of the time series of
$|u(x_0,t)|^2$ (normalized to its maximum). It shows a high peak
of the frequency spectrum near the value $\Omega=1$, in agreement
with the results of the main Theorem \ref{theorem-main}.

{\bf Acknowledgement.} D.P. thanks to W. Craig and V. Konotop for
useful discussions related to the project. D.P. is supported by
the Humboldt and EPSRC fellowships. P.G.K. is supported by NSF
through the grants DMS-0204585, DMS-CAREER, DMS-0505663 and
DMS-0619492.

\appendix
\section{Asymptotic distribution of eigenvalues}

Let us consider the case $\sigma = 1$, when the solution $\phi(x)$
of the ODE (\ref{ODE}) bifurcates to the interval $\mu > 1$ (see
Proposition \ref{lemma-stationary} and Figure \ref{dfig1}). We are
interested in the distribution of eigenvalues of the linear
problem (\ref{spectrum}) as $\mu \to \infty$, assuming that the
solution $\phi(x)$ persists in this limit. It follows from the
scaling transformation below equation (\ref{GP-zero-epsilon}) that
the limit $\mu \to \infty$ of the normalized equation
(\ref{GP-zero-epsilon}) corresponds to the limit $\epsilon \to 0$
in the original GP equation (\ref{GP}). We shall replace $\mu +
\frac{1}{2} = \tilde{\mu}$ and drop tilde notations for the sake
of simplicity. We report here formal results based on asymptotic
methods. Rigorous justification of these results is beyond the
scope of our work.

Denote the ground state of the ODE (\ref{ODE}) by $\phi_0(x)$ such
that $\phi_0(x)$ is even and positive on $x \in \mathbb{R}$ and it
decays to zero as $|x| \to \infty$ sufficiently fast. Using the
substitution $\phi_0(x) = \sqrt{\mu q(\xi)}$ and $\xi =
\frac{x}{\sqrt{2 \mu}}$, we obtain an equation for $q(\xi)$,
\begin{equation}
\label{ODE-WKB} q = 1 - \xi^2 + \frac{1}{4 \mu^2 \sqrt{q}}
\frac{d^2}{d \xi^2} \sqrt{q}, \qquad \forall \xi \in \mathbb{R},
\end{equation}
which is solvable with the nonlinear WKB series \cite{KonKev}. The
main result of the formal WKB theory is that there exists a
classical solution $q_{\mu}(\xi)$ of the ODE (\ref{ODE-WKB}) for
sufficiently large $\mu > 1$ such that
\begin{equation}
\label{ground-state-limit}
\lim_{\mu \to \infty} q_{\mu}(x) = \left\{ \begin{array}{cc} 1 - \xi^2, & \forall |\xi| \leq 1 \\
0, & \forall |\xi| > 1 \end{array} \right.
\end{equation}
The linear problem (\ref{spectrum}) associated with the ground
state $\phi_0(x)$ for $\sigma = 1$ and $\mu + \frac{1}{2} \to \mu$
can be written in variables $v = V(\xi)$, $w = W(\xi)$ and
$\lambda = \mu \Lambda$ for sufficiently large $\mu > 1$. In new
variables, it takes the form
\begin{equation}
\label{linear-problem-WKB} L_+ V = -\Lambda W,  \qquad L_- W =
\Lambda V,
\end{equation}
where
\begin{equation}
L_+ = 3 q(\xi) - 1 + \xi^2 - \frac{1}{4 \mu^2} \frac{d^2}{d
\xi^2}, \quad L_- = q(\xi) - 1 + \xi^2 - \frac{1}{4 \mu^2}
\frac{d^2}{d \xi^2}.
\end{equation}
Eliminating $V(x)$, we close the linear problem
(\ref{linear-problem-WKB}) at the fourth-order ODE
\begin{equation}
L_+ L_- W = \Gamma W, \qquad \Gamma = - \Lambda^2.
\end{equation}
By using the WKB theory (\ref{ground-state-limit}), we consider
the auxiliary eigenvalue problem
\begin{equation}
\label{spectrum-aux} \frac{1}{16 \mu^4} W^{({\rm iv})} - \frac{(1
-\xi^2)}{2 \mu^2} W'' = \Gamma W(\xi), \qquad \forall \xi \in
[-1,1],
\end{equation}
for $W \in L^2([-1,1])$. The entire spectrum of the problem
(\ref{spectrum-aux}) is defined by a set of polynomial solutions
$W = P_m(\xi) = \xi^m + \alpha_{m,m-2} \xi^{m-2} + ... +
\alpha_{m,k} \xi^k$, $\forall m \in \mathbb{N}$, where $k = 1$ if
$m$ is odd and $k = 0$ if $m$ is even. The balance of the largest
term in the ODE (\ref{spectrum-aux}) shows that the eigenvalue
$\Gamma = \Gamma_m$ is found explicitly as $\Gamma_m =
\frac{m(m-1)}{2 \mu^2}$, while all coefficients $\{
\alpha_{m,m-2k} \}_{k = 1}^{[m/2]}$ are uniquely defined.
Converting the values of $\Gamma$ to the values of $\lambda$, we
have found that the linear problem (\ref{spectrum}) associated
with the ground state $\phi_0(x)$ has a set of simple purely
imaginary and symmetric eigenvalue pairs $\{ \pm i \Omega_m \}_{m
\in \mathbb{N}}$, such that
\begin{eqnarray}
\label{eigenvalue-expansion-1-asymptotic}  \lim\limits_{\mu \to
\infty} \Omega_m = \frac{\sqrt{m(m+1)}}{\sqrt{2}}, \qquad \forall
m \in \mathbb{N},
\end{eqnarray}
in addition to the double zero eigenvalue $\lambda = 0$.

Finally, the dark soliton $\phi(x)$ of the ODE (\ref{ODE}) is
obtained asymptotically from the ground state $\phi_0(x)$ by the
factorization $\phi(x) = \phi_0(x) \psi(x)$, where $\psi(x)$ is
odd on $x \in \mathbb{R}$, positive on $x \in \mathbb{R}_+$ and
may approach to the constant values as $|x| \to \infty$
\cite{PFK05}. Using this factorization and the formal asymptotic
analysis, it was shown in \cite{PFK05} that the spectrum of the
linear problem (\ref{spectrum}) associated with the dark soliton
$\phi(x)$ admits a pair of simple purely imaginary eigenvalues
$\pm i \Omega_0$, such that
\begin{eqnarray}
\label{eigenvalue-expansion-3-asymptotic} \lim_{\mu \to \infty}
\Omega_0 = \frac{1}{\sqrt{2}}.
\end{eqnarray}
Although the analysis of \cite{PFK05} was directed to the original GP
equation (\ref{GP}) in the limit of small $\epsilon$ and the
eigenvalue pair was found to be $\tilde{\lambda} \to \pm i
\epsilon$, the scaling transformation to the normalized GP
equation (\ref{GP-zero-epsilon}) implies that $\lambda =
\frac{\tilde{\lambda}}{2^{1/2} \epsilon} \to \pm
\frac{i}{\sqrt{2}}$.

Numerical computations (see Figure \ref{dfig2}) suggests that the
entire spectrum of the linear problem (\ref{spectrum}) associated
with the dark soliton $\phi(x)$ is a superposition between an
infinite set of eigenvalues
(\ref{eigenvalue-expansion-1-asymptotic}) of the linear problem
(\ref{spectrum}) associated with the ground state $\phi_0(x)$ and
the additional pair of eigenvalues
(\ref{eigenvalue-expansion-3-asymptotic}).

Note that the linear eigenmode corresponding to the smallest
eigenvalue $\Omega_0 = \frac{1}{\sqrt{2}}$ may not result in the
periodic solution of the nonlinear PDE system (\ref{PDE}) because
the non-resonance condition $n \neq \sqrt{m(m+1)}$ for all $n,m
\in \mathbb{N}$ is violated in the limit $n,m \to \infty$.
Similarly, the linear eigenmode corresponding to the second
eigenvalue $\Omega_1 = 1$ may not result in the periodic solution
of the PDE system (\ref{PDE}) because the non-resonance condition
$n \neq \frac{\sqrt{m(m+1)}}{\sqrt{2}}$ for all $n,m = 2,3,...$ is
violated at least for $n = 6$ and $m = 8$. In both cases, the
Lyapunov Theorem for persistence of periodic orbit in Hamiltonian
dynamical systems can not be applied \cite{Lyap}.


\begin{thebibliography}{99}

\bibitem{AS} M. Abramowitz and I.A. Stegun, {\em Handbook of
Mathematical Functions with Formulas, Graphs, and Mathematical
Tables} (Dover, New York, 1965), chapter 22.

\bibitem{Adams} R.A. Adams, {\em Sobolev Spaces} (Academic Press
Inc., San Diego, 1978)

\bibitem{Alfimov} G.L. Alfimov and D.A. Zezyulin, "Nonlinear modes
for the Gross--Pitaevskii equation - demonstrative computation
approach", arXiv: nlin.PS/0703006

\bibitem{BG} D. Bambusi and G. Gaeta, "On persistence of invariant
tori and a theorem by Nekhoroshev", Math. Phys. Electr. Journal
{\bf 8}, paper I (2002)

\bibitem{Kon05} V.A. Brazhnyi and V.V. Konotop, "Evolution of a dark soliton
in a parabolic potential: application to Bose--Einstein
condensates", Physical Review A {\bf 68}, 043613 (2003)

\bibitem{Carles} R. Carles, "Remarks on nonlinear Schr\"{o}dinger
equations with harmonic potential", Annales Henri Poincare {\bf
3}, 757--772 (2002)

\bibitem{Freud} G. Freud and G. N\'{e}meth, "On the $L_p$-norms of
orthonormal Hermite functions", Studia Scientiarum Mathematicarum
Hungarica {\bf 8}, 399-404 (1973)

\bibitem{GS} M. Golubitsky and D.G. Schaeffer, {\em
Singularities and Groups in Bifurcation Theory}, vol. 1,
(Springer-Verlag, New York, 1985)

\bibitem{Gus} B.L.G. Jonsson, J. Fr\"{o}hlich, S. Gustafson,  and
I.M. Sigal, "Long time motion of NLS solitary waves in a confining
potential", Annales Henri Poincare {\bf 7}, 621--660 (2006)

\bibitem{Kato} T. Kato, {\em Perturbation theory for linear operators},
(Springer-Verlag, New York, 1976)

\bibitem{Konotop} V.V. Konotop, "Dark solitons in Bose--Einstein condensates: theory"
in {\em "Emergent Nonlinear Phenomena in Bose--Einstein
Condensates"}, Eds. P.G. Kevrekidis, D.J. Franzeskakis, and R.
Carretero--Gonzalez (Springer--Verlag, New York, 2007)

\bibitem{KonKev} V.V. Konotop and P.G. Kevrekidis, "Bohr--Sommerfeld quantization condition
for the Gross--Pitaevskii equation", Physical Review Letters {\bf
91}, 230402 (2003)

\bibitem{Kuksin} S.B. Kuksin, {\em Nearly Integrable
Infinite--Dimensional Hamiltonian Systems} (Springer--Verlag,
Berlin, 1993)

\bibitem{K95} Y.A. Kuznetsov, {\em Elements of Applied Bifurcation
Theory}, 2nd ed., Appl. Math. Sci. {\bf 112} (Springer--Verlag,
New York, 1998)

\bibitem{Lyap} M.A. Lyapunov, {\em The General Problem of the
Stability of Motion} (Taylor and Francis, London, 1992)

\bibitem{PFK05} D.E. Pelinovsky, D. Frantzeskakis, and P.G. Kevrekidis,
"Oscillations of dark solitons in trapped Bose-Einstein
condensates", Physical Review E {\bf 72}, 016615 (2005)

\bibitem{PK07} D.E. Pelinovsky and P.G. Kevrekidis,
"Dark solitons in external potentials", Zeitschrift f\"{u}r
Angewandte Mathematik und Physik, to be published (2007)

\bibitem{proukakis} N.G. Parker, N.P. Proukakis, C.F. Barenghi, and C.S. Adams,
"Dynamical instability of a dark soliton in a
quasi-one-dimensional Bose-Einstein condensate perturbed by an
optical lattice", Journal of Physics B: Atomic Molecular Optical
Physics {\bf 37}, S175--S185 (2004)

\end{thebibliography}
\end{document}